\documentclass[10pt,a4paper,twocolumn,english,pra,aps,showpacs,floatfix,groupedaddress,superscriptaddress]{revtex4-1}

\usepackage{graphicx}
\usepackage{epsfig}
\usepackage[english]{babel}
\usepackage{amsmath}
\usepackage{amssymb}
\usepackage{amsfonts}
\usepackage{longtable}
\usepackage{dcolumn}
\usepackage{bm}
\usepackage{bbm}
\usepackage{color,array}
\usepackage{colortbl}

\begin{document}
\title{Finite temperature line-shapes of hard-core bosons in quantum magnets:\\ A diagrammatic approach tested in one dimension}

\author{Benedikt Fauseweh}
\email{benedikt.fauseweh@tu-dortmund.de}
\affiliation{Lehrstuhl f\"{u}r Theoretische Physik I, Technische Universit\"at Dortmund, Otto-Hahn Stra\ss{}e 4, 44221 Dortmund, Germany}

\author{Joachim Stolze}
\email{joachim.stolze@tu-dortmund.de}
\affiliation{Lehrstuhl f\"{u}r Theoretische Physik I, Technische Universit\"at Dortmund, Otto-Hahn Stra\ss{}e 4, 44221 Dortmund, Germany}

\author{G\"otz S.\ Uhrig}
\email{goetz.uhrig@tu-dortmund.de}
\affiliation{Lehrstuhl f\"{u}r Theoretische Physik I, Technische Universit\"at Dortmund, Otto-Hahn Stra\ss{}e 4, 44221 Dortmund, Germany}

\date{\rm\today}

\begin{abstract}
The dynamics in quantum magnets can often be described by effective models with bosonic excitations obeying a hard-core constraint. Such models can be systematically derived by  renormalization schemes such as continuous unitary transformations or by variational approaches. 
Even in the absence of further interactions the hard-core constraint makes the dynamics of
the hard-core bosons nontrivial. Here we develop a systematic diagrammatic approach to the spectral properties of hard-core bosons at finite temperature. 
Starting from an expansion in the density of thermally excited bosons in a system
with an energy gap, our approach leads to a summation of ladder diagrams.
Conceptually, the approach is not restricted to one dimension, but the one-dimensional case offers the opportunity to gauge the method by comparison to exact results obtained via a mapping to
 Jordan-Wigner fermions. In particular, we present results for the thermal broadening of single-particle spectral functions at finite temperature. The line-shape is found to be asymmetric at elevated temperatures and the band-width of the dispersion narrows with increasing temperature. Additionally, the total number of thermally excited bosons is calculated and compared to 
 various approximations and analytic results. Thereby, a flexible approach is introduced which
 can also be applied to more sophisticated and higher dimensional models.
\end{abstract}


\maketitle

\section{Introduction}

Dynamic correlations generally provide valuable information about the systems under study.
In linear response, for instance, all conceivable susceptibilities are dynamic correlations.
In many spectroscopic experiment{s} dynamic correlations are measured, for instance
dynamic structure factors in scattering experiments or current-current correlations in 
reflectivity or absorption measurements. At low temperatures, such experimental data provide
valuable information about elementary excitations, their mutual interaction, and many
matrix elements.

Theoretically, it is well established that even complex quantum systems can
be described at low energies by simpler effective models. Such models can result for instance from
various renormalization procedures \cite{solyo79,metzn12}. 
Focusing on gapped systems, the elementary excitations
can be viewed generically as quasi-particles the number of which is conserved
which move and interact and
determine the physical properties of the system under study. The corresponding effective
Hamiltonian in terms of these quasi-particles can be derived systematically
by unitary transformations, see for instance Ref.\ \onlinecite{knett03a}, or by variational
approaches, see for instance Ref.\ \onlinecite{haege13a}. Such approaches yield models
in terms of the elementary excitations so that the dynamic response at zero temperature
is captured and an enormous wealth of information is available.
If such models conserve the number of quasi-particles, i.e., the number of excitations,
they are no longer plagued by quantum fluctuations: Their ground state is the vacuum of
the quasi-particles, cf.\ Ref.\ \onlinecite{knett03a}. Thus the only remaining fluctuations
to worry about are thermal ones at finite temperature.

We have two classes for physical realizations of such systems in mind.
Quantum antiferromagnets with a finite spin gap represent one wide class.
In particular various antiferromagnets made from coupled spin dimers
belong to this class, for instance, dimerized spin chains \cite{Knetter00,cavad00,tennant12a}, 
spin ladders \cite{schmi05b,normand11}, two-dimensionally coupled spin ladders \cite{uhrig98c,uhrig04a,nafra11,fisch11a}, and three-dimensionally coupled spin dimers \cite{ruegg05,lake12a}.  The elementary excitations are triplons, i.e.,
$S=1$ hard-core bosons \cite{schmi03c}. The hard-core repulsion comes about 
because the presence of one excitation at a given 
site (dimer) excludes the presence of a second excitation at this site. 
Another example of hard-core bosonic excitations are spin flips in the high-field phase of the
transverse field Ising model. They represent hard-core excitations of a single
flavor {\cite{fause13a}}.

The other class are designed systems made from ultracold bosonic atoms trapped in suitable
optical traps and lattices, for a review see Ref.\ \onlinecite{bloch08}. The setup of
ultracold atoms has the advantage that their number is conserved by construction.
A disadvantage with respect to the considerations presented here is that
there is no scattering process which creates or annihilates an atom
as does inelastic neutron scattering with magnetic triplons or magnons.

The models in terms of conserved hard-core bosons, i.e., their number is a conserved 
quantity, are very successfully analyzed to extract
information at zero temperature, i.e., in the immediate vicinity of the ground state.
But the models are also valid and useful at finite temperatures
unless the temperature is so high that the system enters another phase.


The overall goal of the present article is to benchmark a diagrammatic approach which
allows one to include the thermal fluctuations in a strongly interacting model of
hard-core bosons with the vacuum being the ground state, i.e., 
no bosonic condensate is considered. 
The benchmarking is done for a simple gapped chain model in which only one kind
of hard-core bosons hop from site to site. This model can be treated exactly
by a Jordan-Wigner mapping to free fermions so that a reliable testbed is available.
However, the simple model shares all the basic elements
of more complicated hard-core boson models: Mobile bosons with an infinite local repulsion 
whose number is conserved  so that no quantum fluctuations occur.

We stress that the one dimensional case serves only as a testbed to gauge our method while
the approach is applicable also in higher dimensions and is not restricted to the one dimensional case.
Since only a limited toolset to calculate the finite temperature dynamics of quantum systems in higher dimensions is available, the diagrammatic approach is a promising new technique for a wide range of problems.

By the careful benchmark, we validate an approach which can be applied to all kinds
of particle-conserving gapped models. Such  effective models are no longer
subject to quantum fluctuations. They can be derived systematically by various techniques, 
in particular by continuous unitary transformations \cite{knett03a} so that a large
class of problems can be tackled in this way.
Wide-spread examples are the excitation line-shapes in gapped quantum antiferromagnets
investigated by neutron scattering in one dimension \cite{tennant12a} as well as in three dimensions \cite{ruegg05,lake12a}.
So far the theoretical treatments of these experiments have focused on dispersion relations
$\omega(k)$ and their temperature dependent shifts, neglecting issues of line shape and width.
We highlight the one-dimensional material $[\mathrm{Cu(N}\mathrm{O}_3)_2 \cdot 2.5  (\mathrm{D}_2 \mathrm{O})]$ which is strongly dimerized and exhibits a large gap-to-bandwith ratio so that a particle-conserving  model free of quantum fluctuations is easily derivable \cite{schmi04a}.

The diagrammatic approach to thermal fluctuations is 
systematically controlled by the expansion parameter $\exp(-\beta\Delta)$ where $\beta$ is 
the inverse temperature and $\Delta$ the energy gap.
 We present the direct calculation of the leading order in $\exp(-\beta\Delta)$
of the self energy and the corresponding self-consistent calculation with dressed propagators.
This approach is tested for a model for which numerically exact results are available
for the propagators. We show by comparison of numerical data that both approaches
agree in linear order in $\exp(-\beta\Delta)$. In this comparison we focus on
the position of the single-boson peak in the spectral response, its broadening and
the symmetry of its shape. We show that while the single-boson peak broadens, the overall
bandwidth of the dispersion tends to narrow upon increasing temperature.
This behavior has been observed experimentally \cite{cavad00,ruegg05,tennant12a,lake12a}
and it is at the focus of ongoing theoretical research 
\cite{mikeska06,essler08a,essler08b,essler09a,essler10a}. This underlines its
importance. In view of the constantly improved instrumentation of
inelastic neutron scattering setups a growing interest in reliable theoretical
techniques for the thermal effects on line shapes is to be expected.
We also evaluate the average occupation due to thermal excitations and find
an excellent agreement between the self-consistent diagrammatic result and the exact one.

In view of these results, the self-consistent diagrammatic approach suggests itself for
application to more complicated, extended models 
for gapped quantum antiferromagnets \emph{after} their re-formulation in
terms of conserved quasi-particles to take the quantum  fluctuations into account.
Extensions may comprise longer-range
hoppings, further interactions among the hard-core excitations, and excitations of several flavors.
Also, the application in dimensions greater than one appears promising
because one may generally expect that a diagrammatic perturbative approach works
even better in higher dimensions.

The present article is set up as follows:
In the next section, the considered model is introduced. Second, in Sect.\ III,
we present the diagrammatic approach in detail; a particular focus lies on the
pure cosine band as it results from nearest-neighbor hopping.
In Sect.\ IV we explain how the equivalent fermionic model allows the
numerically exact calculation of the propagators of the hard-core bosons.
In Sect.\ V, we gauge the diagrammatic approach in quantitative comparison
with the exact results and discuss its properties in detail. Finally, Sect.\ 
VI concludes the article.

\section{Model}
\label{sec.model}

In this section we introduce the model and the general quantities, such as the spectral functions, in which we are interested. General observations on the temperature dependence of the spectral weight of hard-core bosons are also presented.

To introduce the diagrammatic expansion for hard-core bosons, we choose a simple 1D model, a single chain of lattice sites. Every site can either be occupied or empty, so that the local Hilbert space is two dimensional. 
In real space the Hamiltonian reads
\begin{align}
\label{eq.Hamiltonian_Local}
H = \sum\limits_i \left(\Delta + \frac{W}{2} \right) b_i^\dagger b_i^{\phantom\dagger} - \sum\limits_i \frac{W}{4} \left( b_i^\dagger b_{i+1}^{\phantom\dagger} + \mathrm{h.c.} \right) ,
\end{align}
where $b_i^\dagger$ and $b_i$ are operators at site $i$, which fulfill the hard-core boson relation,
\begin{align}
\label{eq.hard_core_commutator_local}
\left[ b_j^{\phantom\dagger}, b_i^\dagger \right] &= \delta_{i,j} \left( 1 - 2 b_i^\dagger b_i^{\phantom\dagger} \right).
\end{align}
The Hamiltonian \eqref{eq.Hamiltonian_Local} consists of a local energy term and a nearest neighbour hopping. The energy gap is given by $\Delta>0$, while $W>0$ is the band-width of the dispersion
\begin{align}
\label{eq.dispersion}
\omega(k) = \Delta + \frac{W}{2} \left[ 1-\cos(k) \right].
\end{align}
where $k$ is the momentum of the excitations. The minimum of the dispersion with value $\Delta$ is found at $k=0$.
The ground state of the system is given by the vacuum state.

In typical physical systems, such as spin ladders or spin chains, additional interactions, non-particle-conserving terms as well as longer range hopping are present. We do not consider these terms in our simple model, but we stress that there are a variety of methods to treat such terms which can be combined with our diagrammatic approach. For example, models with non-particle-conserving terms can be mapped to effective, particle-conserving models by continuous unitary transformations (CUT) \cite{Wegner94, GlazekWilson93,*GlazekWilson94, Knetter00,Fischer2010,Krull12}, while additional interactions, besides the hard-core constraint, can be dealt with on a mean-field level. Longer range hopping processes trivially modify the dispersion.

To test our diagrammatic results, we make use of the fact that the Hamiltonian \eqref{eq.Hamiltonian_Local} can be mapped to a system of free fermions by the Jordan-Wigner {t}ransformation \cite{JordanWigner28, ParkinsonFarnell:quantumspinsystems}
\begin{subequations}
\label{eq.Jordan_Wigner}
\begin{eqnarray}
c_j &=& \text{exp}(\pi i \sum_{l<j} b_l^\dagger b_l^{\phantom\dagger}) b_j 
\\
b_j &=& \text{exp}(- \pi i \sum_{l<j} c_l^\dagger c_l^{\phantom\dagger}) c_j 
\end{eqnarray}
\end{subequations}
The operators $c_j^\dagger$ and $c_j$ fulfill fermionic anti-commutator relations.
After a Fourier transformation to reciprocal space, the Hamiltonian \eqref{eq.Hamiltonian_Local} reads
\begin{align}
H = \sum\limits_k \omega(k) c_k^\dagger c_k^{\phantom\dagger} .
\end{align}
In contrast to free bosons, the spectral properties of hard-core bosons at finite temperature are not exactly known. Even though the above fermionic approach can be applied, the calculation of dynamic susceptibilities remains a difficult task 
because of the non-locality of the Jordan-Wigner transformation.  
One example is the single-particle temperature Green function
\begin{align}
G(j, \tau) = - \left\langle T b_j^\dagger(-i\tau) b_0^{\phantom\dagger}(0) \right\rangle ,
\end{align}
where $\tau$  is the imaginary time and $T$ is the time-ordering super operator. The corresponding spectral function reads
\begin{align}
\nonumber
A(p, \omega) &= \frac{-1}{\pi} \lim\limits_{i \omega_\nu \rightarrow \omega + i 0^+} 
\\
& \quad \mathrm{Im} \int\limits_0^\beta \mathrm{d}\tau  e^{i \omega_\nu \tau} \frac{1}{\sqrt{N}} \sum\limits_l e^{-i p l} G(l, \tau) ,
\end{align}
where we introduced the Matsubara frequencies $\omega_\nu$, the inverse temperature $\beta = 1/T${, and  the total number of sites $N$.}

The spectral function is connected to the dynamic structure factor (DSF) by the fluctuation-dissipation theorem
\begin{align}
\label{eq.fluctuation_dissipation}
S(p, \omega) = \frac{1}{1-e^{-\beta\omega}} \left[ A(p, \omega) + A(p, -\omega) \right].
\end{align}
The DSF is the relevant  quantity for many experiments. For instance, it
 is accessible  in inelastic neutron scattering experiments.

Note that there are no contributions from anomalous Green functions,
\begin{align}
G(l, \tau)_\mathrm{anomalous} = - \left\langle T  b_0^\dagger(-i\tau), b_j^{\dagger}(0) \right\rangle ,
\end{align}
since the Hamiltonian \eqref{eq.Hamiltonian_Local} is particle conserving.

Another important difference between hard-core bosons and normal bosons concerns sum rules for spectral functions.
In general the weight of the spectral function is determined by
\begin{align}
 \int\limits_{-\infty}^\infty A(p, \omega) \mathrm{d} \omega = \left\langle \left[ b_p^{\phantom\dagger}, b_p^\dagger \right] \right\rangle.
\end{align}
For normal bosons this is a constant. In contrast, for hard-core bosons, transforming Eq.\ \eqref{eq.hard_core_commutator_local} into Fourier space yields
\begin{align}
\left[ b_p, b_{p'}^\dagger \right] &= \delta_{p,p'} - \frac{2}{N} \sum\limits_{q_3} b_{q_3-p+p'}^\dagger b_{q_3} .
\end{align}
As a result the sum rule reads
\begin{align}
\label{eq.sum_rule}
\int\limits_{-\infty}^\infty A(p, \omega) \mathrm{d} \omega = 1 - 2 n(T)
\end{align}
where $n(T) = \frac{1}{N} \sum_q \langle b_q^\dagger b_q \rangle$ is the thermal occupation. 
At zero temperature, $n(T=0) = 0$ holds and the sum rule yields unity. The opposite case at infinite temperatue 
is $n(T=\infty) = 1/2$; a site is occupied with  probability $50 \%$ and the spectral weight 
over all positive and negative frequencies adds up to zero. 

\section{Diagrammatic Approach}

In this section we introduce the diagrammatic approach to treat the hard-core repulsion at finite temperatures. The key idea is an expansion
in the small parameter $\exp(-\beta\Delta)$. 
In subsection \ref{ssect:gen_calcul} the single-particle self energy is calculated
for arbitrary dispersion relation. In the following subsection \ref{ssect:cos-band} we specialize our results to the pure cosine band and show for this special case, that the shift of the dispersion, due to the finite density of thermal fluctuations, is a second order effect in $\exp(-\beta \Delta)$.

\subsection{General calculations}
\label{ssect:gen_calcul}

Our diagrammatic approach is based on the idea to replace the hard-core boson operators by pure bosonic operators and to enforce the hard-core constraint by an infinite on-site interaction,
\begin{align}
H &\rightarrow H_B = H + H_U  &H_U = \lim\limits_{U \rightarrow \infty} U \sum\limits_{i} b_i^\dagger b_i^\dagger b_i^{\phantom\dagger} b_i^{\phantom\dagger}
\end{align}
This idea {has long been known} under the name of Br\"uckner theory for nuclear matter and He$^3$ \cite{fetterwalecka}.
For low dimensional solid state systems such as spin ladders it {was first proposed}
in Ref.\ \cite{kotov98a}. In contrast to {that investigation}, 
we are not dealing with quantum fluctuations but with thermal fluctuations, i.e., the number of bosons is a constant of motion
in our case. In the original approach anomalous Green functions are present and it is difficult to determine which diagrams 
contribute in leading order in the density of quantum fluctuations. In our application, no quantum fluctuations are 
present. The initial spin Hamiltonian may contain non-particle-conserving terms, but we assume that they
have been eliminated by some renormalizing procedure, for instance by unitary transformations. 
Consequently, we assume that at zero temperature the propagation of the hard-core bosons is described
exactly.
 
Transforming the Hamiltonian $H_B$ into reciprocal space yields
\begin{align}
H_B = \sum\limits_q \omega(q) b_q^\dagger b_q^{\phantom\dagger} + \frac{U}{N} \sum\limits_{q,p,k} b_{q+k}^\dagger b_{-k}^\dagger b_{q+p}^{\phantom\dagger} b_{-p}^{\phantom\dagger}.
\end{align}

Note that in higher dimensions the momenta $q,p,k$ are vectors. 
This does not change the subsequent theoretical calculations. 
In practice, the equations can then be solved using higher dimensional fast Fourier techniques.

\begin{figure}
\centering
\includegraphics[width=1.0\columnwidth]{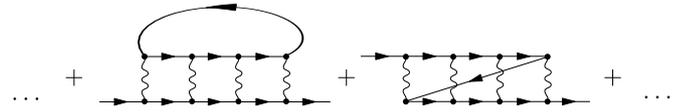}
\caption{Ladder diagrams for the one-particle self energy}
\label{fig.ladder1}
\end{figure}

We expand the single-particle propagator $G(k, \omega)$ in terms of Feynman diagrams. Since the interaction is infinite, a truncated perturbative approach in the interaction strength cannot succeed. 
Instead, we use the density of thermal excitations as small expansion parameter, i.e., we expand
in $\exp(-\beta \Delta)$. The leading order is given by those diagrams that have the fewest number of propagators going backwards
in imaginary time. Thus, in linear order in $\exp(-\beta \Delta)$, we have to sum all diagrams with a single
loop which amounts to the summation of the ladder diagrams, see Fig.\ \ref{fig.ladder1}, following the arguments in Refs.\ \cite{kotov98a, fetterwalecka, Abrikosov:QFT}. We treat the system as a dilute Bose gas. Note that the ladder approximation presented here differs from those in standard textbooks
\cite{fetterwalecka}, where the leading \emph{quantum fluctuations} are
captured which result from the dominant interaction of the quasi-particles with the condensate. 
In contrast to this, our approach captures the leading effects on the spectral function due to
the presence of a small density of  thermally  excited bosons. As a result our approximation breaks down once the density of thermally excited bosons is not small anymore. So the chosen approach represents a low-temperature approximation. 

\begin{figure}
\centering
\includegraphics[width=0.9\columnwidth]{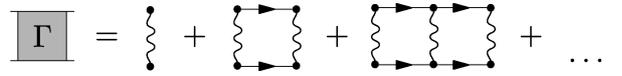}
\caption{Scattering amplitude $\Gamma$}
\label{fig.ladder2}
\end{figure}

\begin{figure}
\centering
\includegraphics[width=0.7\columnwidth]{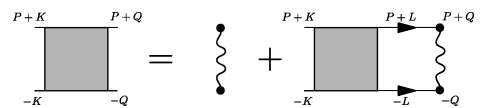}
\caption{Bethe-Salpeter equation for the scattering amplitude}
\label{fig.bethe_salpeter}
\end{figure}

The elementary building block in the ladder diagrams is the scattering amplitude $\Gamma$, defined in Fig.\ \ref{fig.ladder2},
 which can be interpreted as a generalized effective interaction.
It is easily seen that the scattering amplitude fulfills a Dyson-like equation, the Bethe-Salpeter equation, shown in Fig.\ 
\ref{fig.bethe_salpeter}.

To simplify the following expressions we introduce the 2-momentum
\begin{subequations}
\begin{align}
P = \left( p, i \omega_p \right)
\end{align}
and the corresponding summation
\begin{align}
\sum\limits_P = \sum\limits_p \sum\limits_{i \omega_p} .
\end{align}
\end{subequations}
The Bethe-Salpeter equation in Fig.\ \ref{fig.bethe_salpeter} reads in formulae
\begin{equation}
\label{eq.bethe_salpeter}
\begin{aligned}
\Gamma(P) = \frac{U}{N \beta} - \frac{U}{N \beta} \Gamma(P) \sum\limits_L G_0(P+L) G_0(-L).
\end{aligned}
\end{equation}
Note that the scattering amplitude only depends on the total momentum $p$ and not on any relative momenta. 
This is due to the simple structure of $H_U$ in Fourier space.

In the following, we assume that each propagator in the diagrams in Figs.\ \ref{fig.ladder1}, \ref{fig.ladder2} and \ref{fig.bethe_salpeter} is the bare boson propagator
\begin{align}
G_0(P) = \frac{1}{i \omega_p - \omega(p)} .
\end{align}
Below, however, we will relax this assumption in the self-consistent evaluation.
Equation \eqref{eq.bethe_salpeter} can be solved for $\Gamma(P)$ yielding
\begin{align}
\label{eq.gamma_fraction}
\Gamma(P) &=  \frac{U/(N \beta)}{1 + U M(P)},
\end{align}
where
\begin{align}
\label{eq.M_of_p}
M(P) = \frac{1}{N \beta}\sum\limits_L G_0(P+L) G_0(-L).
\end{align}
Since $M(P)$ consists of convolutions of two Green functions, it is of 
order (at least) $\mathcal{O}(1/\omega)$ so that a Hilbert  representation exists; it reads
\begin{align}
\label{eq.spectral_M}
M(\omega, p) = \int\limits_{-\infty}^{\infty} \mathrm{d}x \frac{\rho_p(x)}{\omega - x}
\end{align}
which can be obtained from the imaginary part
\begin{subequations}
\label{eq.rho_p}
\begin{align}
\rho_p(x) &= \frac{-1}{\pi} \lim\limits_{i \omega_p \rightarrow x + i\delta} \mathrm{Im} \frac{1}{N} \sum\limits_{l} \frac{1}{i \omega_p - [\omega(-l) + \omega(p+l)] }  
\nonumber \\ 
&\cdot \left( \frac{1}{e^{- \beta \omega(-l) } - 1} - \frac{1}{e^{\beta \omega(p+l)} - 1} \right) 
\label{eq.rho_p1} \\
		  &= \frac{-1}{N} \sum\limits_{l} \delta(x-[\omega(-l) + \omega(p+l)]) 
		  \nonumber \\ 
		  \label{eq.rho_p2} 
		  &\cdot \left( \frac{1}{e^{\beta \omega(p+l)} - 1} - \frac{1}{e^{- \beta \omega(-l) } - 1} \right) .
\end{align}
\end{subequations}
From the last expression one sees that the spectral function $\rho_p(x)$ is negative so that $M(\omega,p)$ is negative as well for large, positive frequencies.

The spectral density $\rho_p(x)$ can be calculated either analytically or numerically for a given dispersion $\omega(k)$. For computational details we refer the reader to Appendix \ref{app.b}. Note that we already computed the sum over all Matsubara frequencies $\omega_l$ appearing in Eq.\ \eqref{eq.M_of_p} leading to the Bose functions in Eq.\ \eqref{eq.rho_p}. 
The function $\rho_p(x)$ consists of a two-particle continuum describing the spectral properties of two-particle scattering states. For $T \ll \Delta$ the thermal factor can be expanded according to
\begin{align}
\left( \frac{1}{e^{\beta \omega(p+l)} - 1} - \frac{1}{e^{- \beta \omega(-l) } - 1} \right) = 1 + \mathcal{O}\left(e^{-\beta \Delta}\right).
\end{align}
It is not possible to take the limit $U \rightarrow \infty$ already in Eq.\ \eqref{eq.gamma_fraction}, because there exists no spectral representation for $\Gamma (P)$. This is one of the main differences to Ref.\ \cite{sushkov00}, where the Br\"uckner theory was applied to the double-layer Heisenberg antiferromagnet at finite temperatures. In that model, quantum fluctuations are present, but the imaginary part of the self energy was neglected and therefore broadening was omitted. Then, the spectral function $A(p,\omega)$ remains a sharp $\delta$-function even at finite temperature, but its position in frequency depends on temperature.

We know that $M(\omega, p) \propto \mathcal{O}({1}/{\omega})$ for $\omega \rightarrow \infty$. Thus $\frac{U}{1 + U M(\omega, p)} - U \propto \mathcal{O}({1}/{\omega})$ holds. Consequently, there also exists a spectral representation for this quantity
\begin{align} \label{eq.frac_U}
\frac{U}{1 + U M(\omega, p)} - U = \int\limits_{-\infty}^{\infty} \mathrm{d}x \frac{\bar{\rho}_p(x)}{\omega - x},
\end{align}
which again is determined by the imaginary part of the left hand side. We calculate for finite $\omega$ and $U$
\begin{eqnarray}
\nonumber
 &&\frac{-1}{\pi}\lim\limits_{\omega \rightarrow x + i\delta} \mathrm{Im} \left[ \frac{U}{1 + U M(\omega, p)} - U \right] 
 \\
 && \quad =  \frac{ - U^2 \rho_p(x) }{ \left[1 + U \mathcal{P} \int\limits_{-\infty}^{\infty} \frac{\rho_p(y)}{x-y} \mathrm{d}y \right]^2 + \left[ U \rho_p(x) \pi \right]^2 } .
\end{eqnarray}
At this stage, one can take the limit $U \rightarrow \infty$ and define the function
\begin{align}
\label{eq.calc_f}
f_p(x) &= \frac{ -\rho_p(x) }{ \left[\mathcal{P} \int\limits_{-\infty}^{\infty} \frac{\rho_p(y)}{x-y} \mathrm{d}y \right]^2 + \left[ \rho_p(x) \pi \right]^2 }
\end{align}
where $\mathcal{P}$ stands for the {principal} value of the integral.
Again this expression can be evaluated analytically or numerically, depending on the dispersion.
In addition, for large $\omega$, the denominator in \eqref{eq.frac_U} can vanish completely, 
because the real part of $M(\omega, p)$ becomes negative in this region. The vanishing denominator yields an additional $\delta$-function in $\bar{\rho}_p(x)$.

Thus $f_p(x)$ is not the only contribution to $\bar{\rho}_p(x)$ but there also is the signature of an anti-bound state at very high energies in the range $\mathcal{O}{(U)}$. 
The appearance of the anti-bound state in a lattice model with dominant repulsion $U$ between
the elementary excitations is actually to be expected. The propagator of two particles
acquires an additional pole at $U$. In the context of our approach, however, it
is a mathematical artifact because in the final limit $U\to\infty$ the
anti-bound state does not occur directly in any measurable quantity. This would
be different in systems where $U$ is very large, but not infinite, for instance for ultracold atoms in optical lattices. 

We stress that in spite of the limit $U\to\infty$, which makes the anti-bound
state vanish at infinity, it leaves traces at finite energies. To our knowledge,
this has not been discussed in detail before.  We will derive these
effects in the following.
To obtain the position and weight of the anti-bound state we expand the denominator in Eq.\ \eqref{eq.frac_U} for high frequencies $\omega \gg U$ leading to
\begin{align}
\label{eq.expansion_denom}
\frac{U}{1 + U M(\omega, p)} - U \approx \frac{U}{1+U \rho_0(p) \frac{1}{\omega}+U \rho_1(p) \frac{1}{\omega^2}} - U,
\end{align}
where $\rho_0(p)$ and $\rho_1(p)$ are the weight and the first moment of $\rho_p(x)$, respectively, 
\begin{subequations}
\begin{align}
\rho_0(p) = \int\limits_{-\infty}^{\infty} \rho_p(x) \mathrm{d} x, 
\\
\rho_1(p) = \int\limits_{-\infty}^{\infty} x \rho_p(x) \mathrm{d} x .
\end{align}
\end{subequations}
Based on Eq.\ \eqref{eq.rho_p2} one realizes that $\rho_0(p)$ {actually} does not depend on $p$.
Therefore the $p$ dependenc{e} can be ignored in the following calculations.

Extracting the imaginary part from Eq.\ \eqref{eq.expansion_denom} leads to
\begin{subequations}
\begin{align}
\frac{-1}{\pi} \lim\limits_{\omega \rightarrow x + i\delta} \mathrm{Im} \left[ \frac{U \omega^2}{\omega^2 + U \rho_0 \omega + U \rho_1(p)} \right] 
\nonumber \\
 = \frac{-U}{\pi}  \lim\limits_{\omega \rightarrow x + i\delta} \mathrm{Im} \left[ \frac{\omega^2}{(\omega - \omega_1)(\omega - \omega_2)} \right],
\end{align}
where
\begin{align}
\omega_1 &= - \frac{U \rho_0}{2} + \sqrt{\frac{U^2 \rho_0^2}{4}-U\rho_1(p)}, 
\\
\omega_2 &= - \frac{U \rho_0}{2} - \sqrt{\frac{U^2 \rho_0^2}{4}-U\rho_1(p)}. 
\end{align}
\end{subequations}
We point out  that $\omega_1 =\mathcal{O}(U)$ is the energy of the anti-bound state while $\omega_2=\mathcal{O}(U^0)$ 
is a spurious root due to the expansion in $1/\omega$. Now we use Dirac's identity
\begin{align}
 - \pi \delta \left( \omega - \omega_0 \right) = \mathrm{Im} \, \lim\limits_{\delta \rightarrow 0^+} \frac{1}{\omega - \omega_0 + i\delta}
\end{align}
to obtain
\begin{align}
\frac{-1}{\pi} \lim\limits_{\omega \rightarrow x + i\delta} 
\mathrm{Im} \left[ \frac{U \omega^2}{\omega^2 + U \rho_0 \omega + U \rho_1(p)} \right] 
\nonumber \\
= U \left( \frac{x^2}{x-\omega_2} \delta(x-\omega_1) + \frac{x^2}{x-\omega_1} \delta(x-\omega_2) \right) . \label{eq.anti_bond_contribution}
\end{align}
In the following, we can drop the contribution of the spurious root.
Combination with Eq.\ \eqref{eq.calc_f} yields 
\begin{align}
\label{eq.rho_bar}
\bar{\rho}_p(x) = f_p(x)  + U \left( \frac{\omega_1^2}{\omega_1-\omega_2} \delta(x-\omega_1) \right) .
\end{align}

\begin{figure}
\centering
\includegraphics[width=0.7\columnwidth]{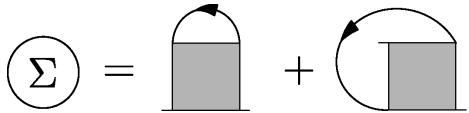}
\caption{Calculation of the self energy}
\label{fig.self_energy}
\end{figure}

Next we address the single-particle self energy. To generate all diagrams in Fig.\ \ref{fig.ladder1} the scattering amplitude must be closed in two ways as  shown in Fig.\ \ref{fig.self_energy}.

If the hard-core bosons can have several flavors, the first diagram has to be counted several times, once for each flavor, because the flavor of the upper boson loop is independent of the one in the lower propagator. 
This is not the case for the second diagram because it does not have an independent boson loop. 
Thus it is counted {only} once even if several flavors are possible.

Due to the simple structure of the interaction, both diagrams yield the same contribution to the self energy
\begin{align}
\Sigma(P) &= \frac{-2}{N \beta} \sum\limits_{K} G_0(K) e^{i \omega_k 0^+} \frac{U}{1 +  U M(P+K)} .
\end{align}
Inserting the spectral representation \eqref{eq.frac_U} yields
\begin{align}
\Sigma(P) &= \frac{-2}{N \beta} \sum\limits_{K} G_0(K) 
\nonumber \\ \label{eq.Sigma_convolved}
          &\cdot e^{i \omega_k 0^+} \left( \int\limits_{-\infty}^{\infty} \mathrm{d}x \frac{\bar{\rho}_{p+k}(x)}{i \omega_p + i \omega_k - x} + U \right) .
\end{align}

Next we sum over all Matsubara frequencies $\omega_k$ and take the limit $U \rightarrow \infty$.  Splitting the self energy into real and imaginary part{s} leads to
\begin{subequations}
\begin{align}
\mathrm{Re} \Sigma(p, \omega) &= \frac{2}{N} \sum\limits_k \left[\frac{\omega}{\rho_0} - \frac{\rho_1(p+k)}{\rho_0^2} + \frac{\omega(k)}{\rho_0} \right] \frac{1}{e^{\beta \omega(k)} - 1} 
\nonumber \\
\label{eq.sigma_real}
					&+  \mathcal{P} \int\limits_{-\infty}^{\infty} \frac{\rho_{\Sigma,p}(x)}{\omega-x} \mathrm{d}x 
					\\
\label{eq.sigma_imag}
\mathrm{Im} \Sigma(p, \omega) &= - \pi \rho_{\Sigma,p}(\omega),
\end{align}
where $\rho_{\Sigma,p}(\omega)$ is the spectral function for the self energy
\begin{align}
\rho_{\Sigma,p} (\omega) &= \frac{2}{N} \sum\limits_{k} f_{p+k}(\omega+\omega(k)) \left[ \frac{1}{e^{\beta \omega(k)} - 1} \right. 
\nonumber \\ 
\label{eq.rho_sigma}
 &- \left. \frac{1}{e^{\beta (\omega + \omega(k))} - 1 } \right] .
\end{align}
\end{subequations}
We draw the reader's attention to 
 the additional contributions in the real part of the self energy \emph{besides} the principal value integral over the spectral function $\rho_{\Sigma,p}(\omega)$. These additional terms stem from the contribution of the anti-bound state in Eq.\ \eqref{eq.anti_bond_contribution}. They are subtle in nature because they represent the left-overs at finite frequencies
 $\omega=\mathcal{O}(U^0)$ of the anti-bound state in \eqref{eq.rho_bar} which itself tends to infinity upon $U\to\infty$. 
In this way, a frequency dependence remains in the real part of the self energy
 which cannot be traced back to a spectral density at finite frequencies.

Next we can calculate our primary quantity of interest, the spectral function of the propagator
\begin{subequations}
\label{eq.A_final}
\begin{align}
A(p, \omega) &= \frac{-1}{\pi} \mathrm{Im} \lim\limits_{i \omega_p \rightarrow \omega + i\delta} G(i \omega_p) 
\\
	&= \frac{-1}{\pi} \frac{\mathrm{Im} \Sigma(\omega,p)}{\left(\omega - \omega(p) - \mathrm{Re} \Sigma(\omega,p) \right)^2 + \left( \mathrm{Im} \Sigma(\omega, p) \right)^2} .
\end{align}
\end{subequations}
The real part of the self energy describes the shift of the peak position due to the interaction with the thermally populated background. We will see that a narrowing of the dispersion will ensue.
The imaginary part describes the broadening of lines of the single excitations due to the hard-core interactions. If the imaginary part of the self energy has only a negligible dependence on $\omega$, the spectral function $A(p, \omega)$ is a symmetric Lorentzian with full width $2 \mathrm{Im} \Sigma$ at half maximum. Below, we will see that this approximation does not hold at elevated temperatures.

One way to improve the results obtained from the ladder diagrams is to calculate the spectral function self-consistently.
Thereby, we realize a conserving approximation in the sense of Baym and Kadanoff \cite{baym61,*baym62}.
In order to do so, each bare propagator must be replaced by the fully dressed propagator
\begin{align}
G(P) = \int_{-\infty}^{\infty} \frac{A(p,x)}{i \omega_p - x} \mathrm{d}x ,
\end{align}
which modifies the resulting equations slightly. On the diagrammatic level, this means that also a number of higher order diagrams are taken into account which consist of propagators with self energy insertions. 
We refer the reader to Appendix \ref{app.a} for more details on the explicit self-consistent calculation.

\subsection{Special case: Cosine band}
\label{ssect:cos-band}

The leading, linear contribution in $\exp(-\beta\Delta)$ in the above general calculations can be evaluated analytically for the cosine band corresponding to nearest-neighbor hopping. This allows us to evaluate the real part of the self energy in this order. For $\rho_p(x)$ we obtain
\begin{eqnarray}
\nonumber
\rho_p(x) &=& -\frac{1}{\pi} \left[ \left[W \cos\left( \frac{p}{2} \right)\right]^2 - \left[2\Delta + W - x\right]^2 \right]^{-1/2}
\\ 
&&+ \mathcal O\left(e^{-\beta \Delta}\right) .
\end{eqnarray}
The corresponding real part vanishes if $\omega$ lies within the band. Using the expansion of $\rho_p(x)$ to calculate the expansion of $f_p(x)$ leads to
\begin{align}
 f_p(x) &= \frac{1}{\pi} \sqrt{\left(W \cos\left( \frac{p}{2} \right)\right)^2 - (2\Delta + W - x)^2 }  
 \nonumber \\ 
 &+ \mathcal O\left(e^{-\beta \Delta}\right) .
\end{align}
For $\rho_{\Sigma,p}(x)$ we deduce
\begin{subequations}
\begin{align}
\rho_{\Sigma,p}(x) = \frac{2}{N} \sum\limits_k f_{p+k}(x+\omega(k)) e^{-\beta\omega(k)} + \mathcal O\left(e^{-2\beta \Delta}\right) ,
\end{align}
where
\begin{align}
\frac{1}{e^{\beta \omega(k)}-1} - \frac{1}{e^{\beta (x+\omega(k))}-1} = e^{-\beta\omega(k)} + \mathcal O\left(e^{-2\beta \Delta}\right),
\end{align}
\end{subequations}
is used for $x \gtrsim \Delta$.
We aim at the real part of the self energy
describing the shift of the spectral function $A(p, \omega)$. The expansion in $\exp(- \beta \Delta)$ yields
\begin{align}
\mathrm{Re} \, \Sigma(p, \omega) &= \frac{2}{N} \sum\limits_k \left[-\omega - \rho_1(p+k) -\omega(k) \phantom{\int\limits_{-\infty}^{\infty} }\right.  
\nonumber \\
&+ \left. \mathcal{P} \int\limits_{-\infty}^{\infty} \frac{f_{p+k}(x+\omega(k))}{\omega-x} \mathrm{d}x \right] e^{-\beta \omega(k)} 
\nonumber \\ 
&+ \mathcal O\left(e^{-2\beta \Delta}\right) \label{eq.real_part_analytics} ,
\end{align}
where we made use of 
\begin{align}
\rho_0 = -1 + \mathcal O\left(e^{-\beta \Delta}\right) .
\end{align}
Next we consider $\rho_1(p+k)$ and expand it in $\exp(-\beta \Delta)$
\begin{subequations}
\label{eq.rho_1_expansion}
\begin{align}
\rho_1(p) &= \int\limits_{-\infty}^\infty \rho_p(x) x \mathrm{d}x , 
\\
	   &= -2 \bar{\omega} + \mathcal O\left(e^{-\beta \Delta}\right) .
\end{align}
\end{subequations}
where $\bar{\omega}$ is the average value of the dispersion in the Brillouin zone. 
For the cosine band this simply is 
\begin{align}
\label{eq.omega_bar}
\bar{\omega} = \Delta + \frac{W}{2}.
\end{align}
Finally, we compute the the principal value integral in \eqref{eq.real_part_analytics} for the cosine band
to reach
\begin{align}
&\mathcal{P} \int\limits_{-\infty}^{\infty} \frac{f_{p+k}(x+\omega(k))}{\omega-x} \mathrm{d}x 
\nonumber \\
&= \omega - \left[\Delta + \frac{W}{2}(1+\cos(k))\right] + \mathcal O\left(e^{-\beta \Delta}\right) \label{eq.principal_value_expansion} , 
\end{align}
as long as $|\omega - \left[\Delta + \frac{W}{2}(1+\cos(k))\right]| < |W \cos(p/2+k/2)|$ holds. This is always true for $\omega = \omega(p)$.
Inserting Eqs. \eqref{eq.rho_1_expansion}, \eqref{eq.rho_bar}, and \eqref{eq.principal_value_expansion} into the real part of the self energy \eqref{eq.real_part_analytics} yields
\begin{equation}
\begin{aligned}
\mathrm{Re} \, \Sigma(p, \omega) 
&= 0 + \mathcal O\left(e^{-2\beta \Delta}\right) .
\end{aligned}
\end{equation}

Since our diagrammatic expansion is correct in first order in $\exp(-\beta \Delta)$ we conclude that the first order corrections in $\exp(-\beta \Delta)$ to the real part of the self energy vanish rigorously. 
We will confirm this conclusion by evaluating the equivalent fermionic model in the next sections.
Hence the shift of the dispersion due to finite temperature for cosine bands is a second order effect 
$\propto \exp(-2\beta \Delta)$. Interestingly, this result in one
dimension also holds in the case of multi-flavor hard-core bosons such as
triplons because the multiplicity of the flavors  only affects the prefactor of the self energy.

\section{Equivalent fermionic model}
\label{sec:fermionization}

In this section an equivalent fermionic model is introduced by means of the Jordan-Wigner transformation. This mapping allows us to calculate the finite temperature dynamics numerically
based on the evaluation of Pfaffians. This approach is limited to a certain class of one dimensional systems but allows us to gauge the diagrammatic approach.

The Hamiltonian \eqref{eq.Hamiltonian_Local} can be interpreted as an anisotropic  spin model, the XX chain. Note that $S_i^+=S_i^x + i S_i^y$ is identical to the hard-core boson creation
operator $b_i^{\dag}$, see Eq.\ \eqref{eq.hard_core_commutator_local}. The open-ended $N$-site $S=1/2$ XX chain in a homogeneous magnetic field $h$ is defined by
\begin{equation} 
  \label{eq:2.1}
  H_\text{XX}= -J \sum_{i=1}^{N-1} (S_i^x S_{i+1}^x +S_i^y S_{i+1}^y) +h
  \sum_{i=1}^N (S_i^z+ \frac 12).
\end{equation}

$H_\text{XX}$ is one of the simplest quantum many-body systems because many of its
properties can be derived from those of non-interacting lattice
fermions. Nevertheless it shows non-trivial dynamics.

$H_\text{XX}$ can be
mapped \cite{LSM61,Kat62} to a Hamiltonian of noninteracting fermions, 
\begin{equation}
  \label{eq:2.2}
  H_\text{F} =  -\frac{J}{2} \sum_{i=1}^{N-1}  (c_i^{\dag} c_{i+1} +
  c_{i+1}^{\dag} c_i) 
+ h \sum_{i=1}^N c_i^{\dag} c_i 
\end{equation}
by means of the Jordan-Wigner transformation \eqref{eq.Jordan_Wigner}. 
The diagonal form of the fermion Hamiltonian is
\begin{equation}
  \label{eq:2.5}
  H_\text{F} = \sum_k \omega(k) c_k^{\dag} c_k 
\end{equation}
where the operators $c_k^{\dag}$ and $ c_k$ create and destroy a
fermion in a one-particle eigenstate, respectively.  The
one-particle energy eigenvalues
are
\begin{equation}
  \label{eq:2.6}
  \omega(k) = -J \cos k+h , k= \frac{\nu \pi}{N+1}
, \nu= 1, \cdots,N
\end{equation}
and the eigenvectors are sinusoidal functions of the site index $i$. 
Obviously the dispersion relations (\ref{eq:2.6}) and \eqref{eq.dispersion} are
identical for $J=\frac W2$ and $h=\Delta + \frac W2$.

In the fermionic ground state all single-particle states with negative energies are
occupied  while all other states are empty. For $|h|>J$  the
ground state is either completely occupied
or completely empty. In the intermediate field range, $|h| <
J$, the ground state contains a partially filled band of
Jordan-Wigner fermions.

We are interested in the
correlation functions $\langle S_i^x (t)S_j^x \rangle$ which have a rather
complicated structure in the fermionic representation.
With the fermionic identity
\begin{equation}
\exp( i\pi {  c_l^{\dag} c_l }) = (c_l^{\dag} + c_l)(c_l^{\dag} - c_l)
\label{II.6}
\end{equation}
applied to the Jordan-Wigner transformation \eqref{eq.Jordan_Wigner}, this correlation function may be expressed
in terms of the auxiliary operators $A_l = c_l^{\dag} + c_l$ and
$B_l = c_l^{\dag} - c_l$ as follows:
\begin{widetext}
\begin{equation}
\langle S_i^x (t) S_j^x\rangle = \frac{1}{4}
\langle A_1(t) B_1(t) A_2(t) B_2(t) ... A_{i-1}(t)
B_{i-1}(t) A_i(t) A_1 B_1 A_2 B_2 ...A_{j-1} B_{j-1} A_j\rangle \, .
\label{II.7}
\end{equation}
This expectation value of a product of $2(i+j-1)$
fermion operators may be expanded in terms of
two-point expectation values using Wick's theorem  \cite{Gau60}.
The result is most compactly expressed as a Pfaffian: 
\begin{displaymath}
4 \langle S_i^x(t) S_j^x\rangle \; \;  =   \; \; \; \; \; \; \; \; \; \; \; \;
\; \; \; \; \; \; \; \; \; \; \; \; \; \; \; \; \; \; \; \; \; \; \; \;
\; \; \; \; \; \; \; \; \; \; \; \; \; \; \; \; \; \; \; \; \; \; \; \;
\; \; \; \; \; \; \; \; \; \; \; \; \; \; \; \; \; \; \; \; \; \; \; \;
\end{displaymath}
\begin{equation}
 \left. \begin{array}{ccccccc}
|\langle A_1(t)B_1(t)\rangle&\langle A_1(t)A_2(t)\rangle& \cdots &
\langle A_1(t)A_1\rangle & \langle A_1(t)B_1\rangle&
 \cdots & \langle A_1(t)A_j\rangle \\
        &\langle B_1(t)A_2(t)\rangle& \cdots &
        \langle B_1(t)A_1\rangle & \langle B_1(t)B_1\rangle & \cdots &
        \langle B_1(t)A_j\rangle \\
   &   & \cdots & \cdots & \cdots & \cdots & \cdots \\
   &   &        & \cdots & \cdots & \cdots & \cdots \\
   &   &        &        & \cdots & \cdots & \cdots \\
   &   &        &        &        & \cdots & \cdots \\
   &   &        &        &        &        & \langle B_{j-1}A_j\rangle \\
   \end{array}
   \right| \, .
\label{II.8}
\end{equation}
The square of the Pfaffian is equal to the determinant of the antisymmetric
matrix with the elements of (\ref{II.8}) above the diagonal. Other
properties of Pfaffians can be found in  the literature \cite{GH64}.
 The numerical evaluation
of Pfaffians proceeds along similar lines as that of
determinants. Many matrix elements can be reduced to zero by operations which are
known to leave the value of the Pfaffian invariant. After production
of sufficiently many zero elements the evaluation of the Pfaffian
becomes trivial due to an expansion theorem. An implementation along
these lines was described by Derzhko and Krokhmalskii \cite{DK98}. We
use a similar algorithm here. An alternative
recursive scheme for evaluating Pfaffians was used by Jia and
Chakravarty \cite{JC06}.

The matrix elements of (\ref{II.8}) can be evaluated from the expressions \cite{CG81}

\begin{subequations}
\begin{eqnarray}
\langle A_j(t) A_l\rangle &=&
\frac{2}{N+1} \sum_k \sin kj \sin kl [ \cos \omega(k)
t - i \sin \omega(k) t \tanh \frac{\beta \omega(k)}{2}]
\label{II.9}
\\
\langle A_j(t) B_l\rangle &=&
\frac{2}{N+1} \sum_k \sin kj \sin kl [ i \sin \omega(k)
t - \cos \omega(k) t \tanh \frac{\beta \omega(k)}{2}] \, .
\label{II.10}
\end{eqnarray}
\end{subequations}
\end{widetext}
along with the relations
\begin{subequations}
\begin{eqnarray}
\langle B_j(t)B_l\rangle =
-\langle A_j(t)A_l\rangle 
\\
 \langle B_j(t)A_l\rangle =
-\langle A_j(t) B_l\rangle \, .
\label{II.11}
\end{eqnarray}
\end{subequations}
All elements of type (\ref{II.9}) with odd $j-l$ and
all elements of type (\ref{II.10}) with even $j-l$ vanish.
In fact, for $t=0$ the elements (\ref{II.9}) are zero for all $j \neq l$.

A word on boundary conditions is in order at this point. The mapping
between the spin and fermion Hamiltonians as given above is only
possible for open boundary conditions. For cyclic boundary conditions
the Jordan-Wigner transformation generates different Hamiltonians
for even and odd total fermion numbers, respectively. This makes the
calculation of the dynamic correlations $\langle S_i^x (t)S_j^x
\rangle$  extremely awkward, if not impossible, since every $S^x$
operator switches back and forth between subspaces of even and odd
fermion numbers. We therefore stick to open boundary conditions.

In order to make sure that open-chain numerical
results pertain to the thermodynamic limit, only spins sufficiently
far from the boundaries of sufficiently long chains may be
considered. Then the finite group
velocity of the Jordan-Wigner fermions  prevents the occurrence of
``echoes'' reflected from the chain boundaries in the dynamic
correlations for short enough times.

The quantity of interest in this study is the DSF $S^x(q,\omega)$, since it is directly related to the spectral function $A(q,\omega)$ by the fluctuation-dissipation theorem.
$S^x(q,\omega)$  is  the Fourier transform with respect to
space and time of the dynamic correlation $\langle S_i^x (t)S_j^x
\rangle$
\begin{equation}
  \label{eq:dsf}
  S^x(q,\omega)=\sum_n \int_{-\infty}^{\infty} dt \, \exp
  [-i(qn-\omega t)] \langle S_l^x (t)S_{l+n}^x \rangle.
\end{equation}

$S^x(q,\omega)$ is determined using fast Fourier transform
algorithms. A technical problem occurs at low temperature, where  
$\langle S_i^x (t)S_j^x \rangle$ displays slow power-law asymptotics
at long times leading to spurious oscillations in
$S^x(q,\omega)$ if the Fourier transform is performed using a finite
time interval. However, for the purposes of the present study only
the position and width of the dominant peak (or rather ridge) in
$S^x(q,\omega)$ is relevant so that no filtering or asymptotic
continuation techniques had to be applied.

All results discussed here
are for the gapped case, $|h|>J$ (or $\Delta>0$). 
DSF data for the case $|h|<J$ were published some years ago \cite{my38} as
were results for dimerized XX chains \cite{my40}.

\section{Results}
\label{sec:results}

In this section we discuss the results of the diagrammatic approach in detail. First we compare the line-shapes calculated by the diagrammatic approach with the numerically exact results obtained 
from the equivalent fermionic model in subsection \ref{ssec:comparison}. Here we also verify our 
analytical finding concerning the shift of the peak position from subsection \ref{ssect:cos-band}. Next 
we extract the thermal occupation function from the single-particle Green function and compare it to various approximations and to the exact result. In subsection \ref{ssec:details_diagrammatic}, we study the real and imaginary parts of the self energy obtained by the diagrammatic approach and discuss their general features. In the final two subsections \ref{ssec:broadening} and \ref{ssec:band_narrowing} we show the finite temperature broadening over a wide range of temperatures for the 
two modes defining the minimum and the maximum of the single-boson dispersion, respectively.
We also discuss how the single-particle band narrows in energy at finite temperature.

\subsection{Comparison to exact fermionic evaluation}
\label{ssec:comparison}

\begin{figure}
\centering
\includegraphics[width=1.0\columnwidth]{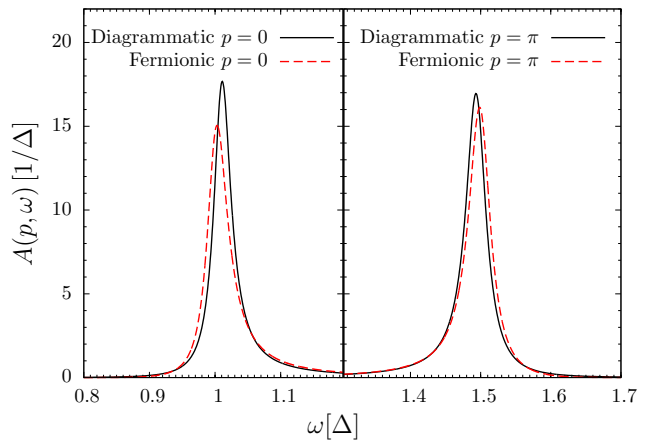}
\caption{(Color online) Comparison between the spectral function obtained from the fermionic approach and from the diagrammatic expansion. Parameters are $p=0$ and $p=\pi$ with $W=0.5\Delta$ and $T=0.434\Delta$, i.e., $\exp(-\beta\Delta)=0.100$}.
\label{fig.Joachim_compare1_W05}
\end{figure}

\begin{figure}
\centering
\includegraphics[width=1.0\columnwidth]{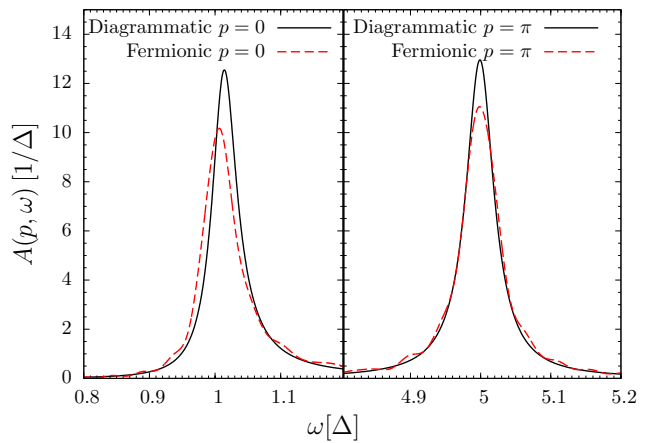}
\caption{{(Color online)} Comparison between the spectral function obtained from the fermionic approach
 and from the diagrammatic expansion. Parameters are $p=0$ and $p=\pi$ with $W=4\Delta$ and $T=0.434\Delta$, i.e., $\exp(-\beta\Delta)=0.100$}.
\label{fig.Joachim_compare1_W4}
\end{figure}

Here, we  compare the results of the diagrammatic expansion to the results of the exact fermionic evaluation. 
The latter are exact except for finite size or finite time effects. Thus the fermionic results will serve as testbed for the diagrammatic approach. The quantity of interest is the single particle spectral function $A(p,\omega)$. It is connected to the DSF by Eq.\ \eqref{eq.fluctuation_dissipation}. It turns out that the spectral function has almost no weight for $\omega<0$ in the parameter regime which we are focusing on. Therefore, we can approximate the relation between $A(p,\omega)$ and $S(p,\omega)$ by
\begin{align}
 S(p, \omega) = \frac{1}{1-e^{-\beta\omega}} A(p, \omega) \, \, , \, \, \omega>0
\end{align}

All diagrammatic results presented are calculated self-consistently, if not denoted otherwise.
In Fig.\ \ref{fig.Joachim_compare1_W05} we compare the spectral function at finite temperature 
$T=0.434\Delta$, i.e., $\exp(-\beta\Delta)=0.100$ obtained by the fermionic approach and by diagrammatic expansion for a narrow band-width $W=0.5\Delta$. The results agree very well for momentum $p=\pi$, while a slight difference appears for $p=0$. To understand the main effects qualitatively, the following argument helps. The weight of the peak changes only very little
as function of temperature, see for instance Figs.\ \ref{fig.Joachim_compare4_W05} and \ref{fig.Joachim_compare4_W4} below. Thus the height of the peak is inversely proportional
to its width. The width is proportional to the imaginary part of the self energy, see Eq.\ \eqref{eq.A_final}, which in turn is exponentially small, namely proportional to $\exp(-\beta\Delta)$.
Thus, the height is proportional to $\exp(\beta\Delta)$. This very strong dependence
on the temperature ($\beta=1/T$) puts the discrepancy between the height of the two curves for $p=0$
into perspective. If $\exp(\beta\Delta)$ is large, let us say $\exp(\beta\Delta)=10$,
a slight inaccuracy of $2\%$ in the gap, which is modified by the real part of
the self energy, induces a $5\%$ error in the height. For $\exp(\beta\Delta)=100$
the relative error would even rise to $10\%$.

The shift of the peak position at finite temperature seems to be overestimated by the diagrammatic expansion. Since the shift is a second order effect $\exp(-2\beta \Delta)$ and the diagrammatic expansion is only correct in first order $\exp(-\beta \Delta)$ such deviations
can be expected. We attribute this difference to two sources: (i) The ladder approximation is not able to capture all relevant physical processes at this fairly elevated temperature. (ii) The data obtained in the fermionic approach has a finite resolution in the time domain which implies 
some inaccuracies in the frequency domain. 
We do not include peaks at lower temperatures because they become
very quickly extremely sharp so that it is difficult to evaluate their shape with appropriate numerical
precision. This holds true for the numerical evaluation of both 
the exact and of the diagrammatic approach.
We emphasize, however, that by construction the diagrammatic approach becomes better and better for lower and lower temperature.

Figure \ref{fig.Joachim_compare1_W4} shows the spectral functions for the wide band case $W=4\Delta$ at finite temperature $T=0.434\Delta$, i.e., $\exp(-\beta\Delta)=0.100$. Due to the larger band-width, the group velocity of the excitations is increased and the finite system size in the fermionic evaluation induces additional errors. This is clearly visible in the additional wiggling in the spectral function obtained in the fermionic approach.

\begin{figure}
\centering
\includegraphics[width=1.0\columnwidth]{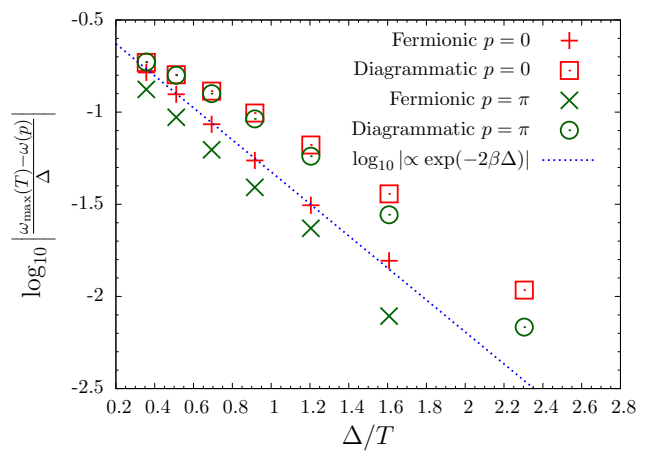}
\caption{{(Color online)} Peak shift for $W=0.5\Delta$ as function of the inverse temperature.  Crosses represent results obtained in the fermionic approach while boxes and circles represent results obtained by the diagrammatic expansion. The function $\log_{10}(A \cdot \exp(-2\beta \Delta)$ has been fitted to the data using the parameter $A$.}
\label{fig.Joachim_compare2_W05}
\end{figure}

\begin{figure}
\centering
\includegraphics[width=1.0\columnwidth]{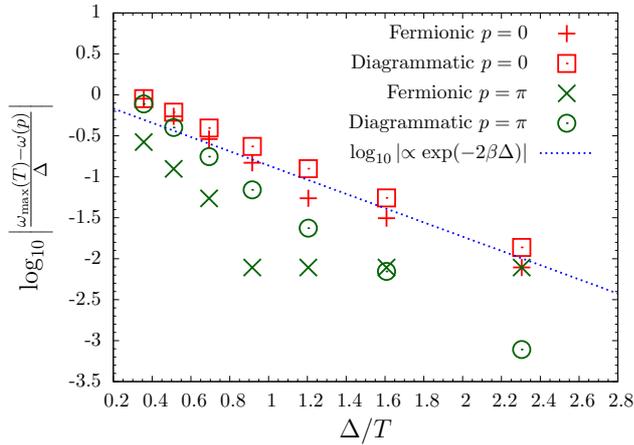}
\caption{{(Color online)} Peak shift for $W=4\Delta$ as function of the inverse temperature. Crosses represent results obtained in the fermionic approach while boxes and circles represent results obtained by the diagrammatic expansion. The function $\log_{10}(A \cdot \exp(-2\beta \Delta)$ has been fitted to the data using the parameter $A$.}
\label{fig.Joachim_compare2_W4}
\end{figure}

Next we want to study the shift of the peak position in more detail.
In Figs.\ \ref{fig.Joachim_compare2_W05} and \ref{fig.Joachim_compare2_W4} the shift for the narrow and for the wide band case is depicted as function of the inverse temperature.
Both methods indicate that the shift is a second order effect in $\exp(-\beta \Delta)$, verifying our analytical argument, see Sec.\ \ref{ssect:cos-band}. The diagrammatic expansion overestimates the shift as we have already seen in Figs.\ \ref{fig.Joachim_compare1_W05} and \ref{fig.Joachim_compare1_W4}.
Note that for $W=4\Delta$ and $p=\pi$ four data points obtained by the fermionic approach show the same shift. This error is caused by the finite frequency resolution used in the fermionic approach. We stress again that the exact fermionic expressions are difficult to evaluate numerically, in particular for low temperatures.

\begin{figure}
\centering
\includegraphics[width=1.0\columnwidth]{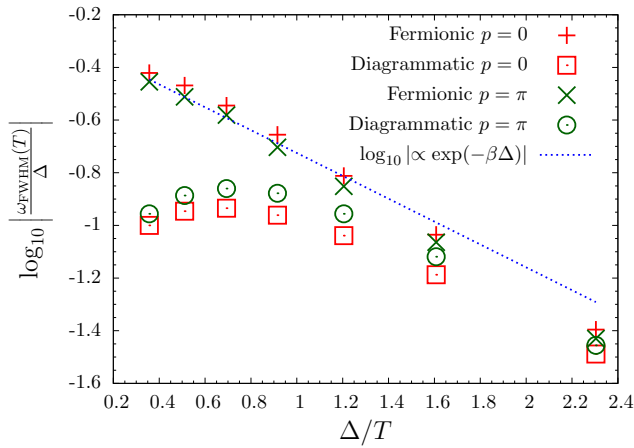}
\caption{{(Color online)} Full width at half maximum of the peak for $W=0.5\Delta$ as function of the inverse temperature. Crosses represent results obtained in the fermionic approach while boxes and circles represent results obtained by the diagrammatic expansion. The function $\log_{10}(A \cdot \exp(-\beta \Delta))$ has been fitted to the data using the parameter $A$.}
\label{fig.Joachim_compare3_W05}
\end{figure}

\begin{figure}
\centering
\includegraphics[width=1.0\columnwidth]{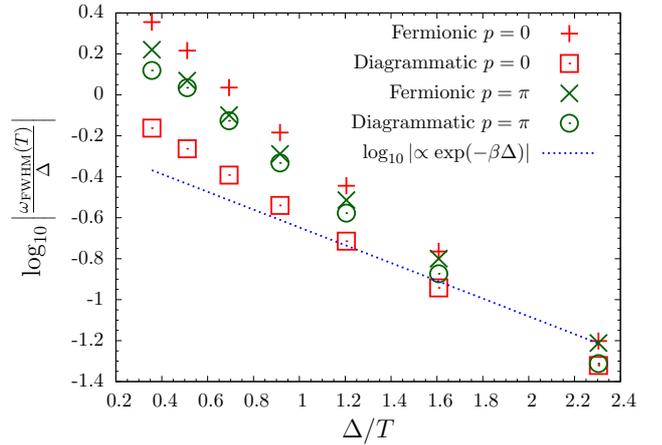}
\caption{{(Color online)} Full width at half maximum of the peak for $W=4\Delta$ as function of the inverse temperature. Crosses represent results obtained in the fermionic approach while boxes and circles represent results obtained by the diagrammatic expansion. The function $\log_{10}(A \cdot \exp(-\beta \Delta))$ has been fitted to the data using the parameter $A$.}
\label{fig.Joachim_compare3_W4}
\end{figure}

The width of the spectral functions is also investigated. It is measured by the full width at half maximum. The data is depicted in Figs.\ \ref{fig.Joachim_compare3_W05} and \ref{fig.Joachim_compare3_W4}. For low temperatures, 
the data of both approaches agrees well. Both data sets support our analytic finding in Eq.\ \eqref{eq.sigma_imag} that the 
 width is a first order effect in $\exp(-\beta \Delta)$. 
 Upon increasing temperature the diagrammatic expansion underestimates the broadening of the line-shape. This can be attributed to the missing diagrams $\propto \exp(-2\beta \Delta)$ not included in our approach. These diagrams describe additional scattering processes increasing the decoherence and broadening the line-shape further. For $W=0.5\Delta$ and very high temperatures we even see that the broadening obtained from the ladder approximation decreases, which clearly indicates that the ladder diagrams  no longer capture the dominant scattering processes.
 
Finally, we consider the temperature dependence of the weight of the spectral function. More specifically, we plot the 
deviation from unity in logarithmic scale $\log_{10}(1 - \int_{-\infty}^\infty A(p, \omega))$ for the narrow band in Fig.\ \ref{fig.Joachim_compare4_W05} and for the wide band in Fig.\ \ref{fig.Joachim_compare4_W4} versus the inverse temperature. For both cases an amazing agreement between the two methods is found. Since we have already shown in Sect.\ \ref{sec.model} that the weight of the spectral function is directly connected to the thermal occupation, see Eq.\ \eqref{eq.sum_rule}, we expect that the thermal occupation is captured very accurately by the diagrammatic expansion
in agreement with the exact fermionic results, see also next subsection. 

\begin{figure}
\centering
\includegraphics[width=1.0\columnwidth]{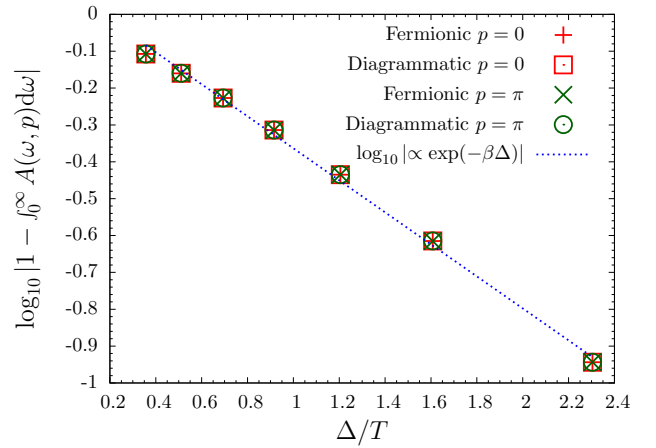}
\caption{{(Color online)} Weight of the spectral function for $W=0.5\Delta$ as function of the inverse temperature. Crosses represent results obtained in the fermionic approach while boxes and circles represent results obtained by the diagrammatic expansion. The function $\log_{10}(A \cdot \exp(-\beta \Delta))$ has been fitted to the data using the parameter $A$.}
\label{fig.Joachim_compare4_W05}
\end{figure}

\begin{figure}
\centering
\includegraphics[width=1.0\columnwidth]{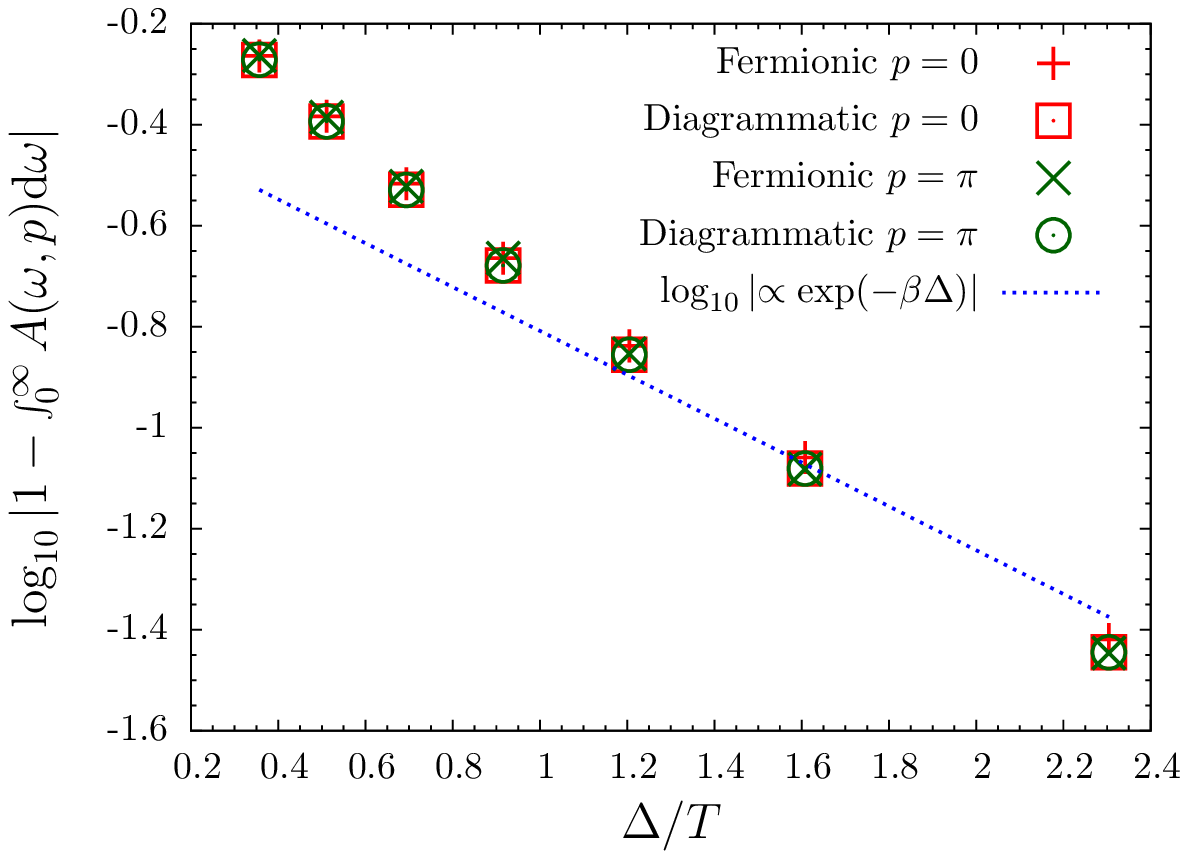}
\caption{{(Color online)} Weight of the spectral function for $W=4\Delta$ as function of the inverse temperature. Crosses represent results obtained in the fermionic approach while boxes and circles represent results obtained by the diagrammatic expansion. The function $\log_{10}(A \cdot \exp(-\beta \Delta))$ has been fitted to the data using the parameter $A$.}
\label{fig.Joachim_compare4_W4}
\end{figure}

\subsection{Thermal occupation}

Here we focus on the thermal occupation which can be determined  from the spectral function by evaluating the integral
\begin{align}
n(T) = \frac{1}{2\pi} \int\limits_0^{2\pi} \langle b_k^\dagger b_k \rangle \mathrm{d}k =  \int\limits_0^{2\pi} \int\limits_{-\infty}^\infty \frac{A(k,\omega)}{e^{\beta \omega}-1} \mathrm{d} \omega \mathrm{d}k.
\end{align}
Since the Jordan-Wigner Transformation maps the hard-core bosons to fermions without interaction, the exact expression for the thermal occupation is easily available
\begin{subequations}
\begin{align}
n(T) &= \frac{1}{2\pi} \int\limits_0^{2\pi} \langle b_k^\dagger b_k \rangle \mathrm{d}k = \frac{1}{N} \sum\limits_i \langle b_i^\dagger b_i \rangle  
\\
										&= \frac{1}{N} \sum\limits_i \langle c_i^\dagger c_i \rangle 
										= \frac{1}{2\pi} \int\limits_0^{2\pi} \frac{1}{e^{\beta \omega(k)} +1} \mathrm{d}k .
\end{align}
\end{subequations}
It is not possible to calculate the momentum dependent thermal occupation analytically, because it includes many-particle correlation functions in the fermionic picture, see Sec.\ \ref{sec:fermionization}.

\begin{figure}
\centering
\includegraphics[width=1.0\columnwidth]{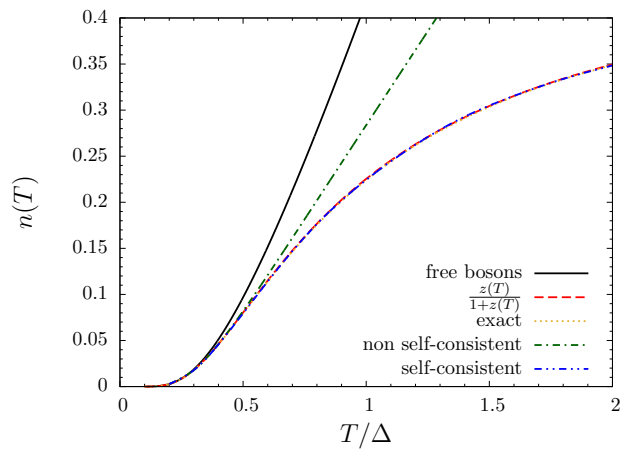}
\caption{{(Color online)} Temperature dependence of the occupation function for $W=0.5\Delta$. Comparison between a simple bosonic approximation, the non-self-consistent solution, the self-consistent solution, the approximate statistics from Ref.\ \cite{troyer94}, and the exact fermionic expression. }
\label{fig.thermal_occupation}
\end{figure}

A crude estimate for the occupation function of hard-core bosons was proposed in Ref.\ \cite{troyer94},
\begin{align}
 n(T,k)_\mathrm{app} = \frac{e^{-\beta \omega(k)}}{1+z(T)},
\end{align}
where $z(T) = \sum_k \exp(- \beta \omega(k))$. It is correct for flat bands $W=0$, but often used as a first approximation also in the case of dispersive bands, see for instance
 Refs.\ \onlinecite{normand11, exius10}. The implied approximate thermal occupation reads
\begin{align}
 n(T)_\mathrm{app} = \frac{z(T)}{1+z(T)}.
\end{align}
We call the thermal occupation approximate hard-core statistics below because it
only captures the local aspects of repelling bosons.
The approximate statistics captures the correct values for $T=0$ and $T=\infty$ for nonzero band-width. We therefore expect deviations from the exact expression to appear for moderate temperatures $T \sim \Delta$ and wide bands $W \gg \Delta$.
The thermal occupation for the narrow band $W=0.5\Delta$ is depicted in Fig.\ \ref{fig.thermal_occupation}. 
For low $T$ the precise statistics does not matter, so that even the free boson statistics
\begin{align}
 n(T)_\mathrm{boson} = \frac{1}{2\pi} \int\limits_0^{2\pi} \frac{1}{e^{\beta \omega(k)} - 1} \mathrm{d}k
\end{align}
captures the correct behaviour.
 This changes distinctively at higher temperatures. The occupation number for free bosons has no bound for rising temperature, while for hard-core bosons the limit $n(T=\infty) = 1/2$ has to be fulfilled. Therefore the free boson approximation breaks down at $T \gtrsim 0.3 \Delta$. The non-self-consistent calculation improves the statistics beyond this point, but also breaks down once temperature reaches $T \gtrsim 0.6 \Delta$. Fortunately, the self-consistent calculation stays very close to the exact result, even for temperatures far above the energy gap. For the narrow band the approximate statistics $n(T)_\mathrm{app}$ is indistinguishable from the exact curve.
 
\begin{figure}
\centering
\includegraphics[width=1.0\columnwidth]{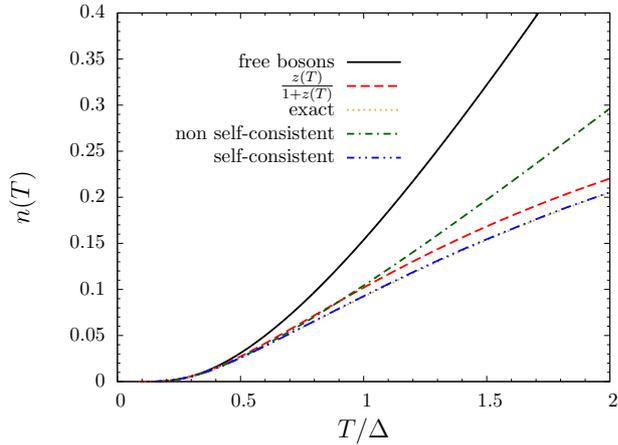}
\caption{{(Color online)} Temperature dependence of the occupation function for $W=4\Delta$. Comparison between a simple bosonic approximation, the non-self-consistent solution, the self-consistent solution, the approximate
 statistics from Ref.\ \cite{troyer94}, and the exact fermionic expression.}
\label{fig.thermal_occupation_W4}
\end{figure}

Figure \ref{fig.thermal_occupation_W4} shows the thermal occupation for the wide band $W=4\Delta$ for the same temperature range as in the narrow band case. The thermal occupation does not grow as fast as for the narrow
band due to the larger energy scale $W$. The free boson approximation holds up to $T \approx 0.4 \Delta$ while the non-self-consistent calculation is correct up to $T \approx 0.8 \Delta$. Again the self-consistent solution agrees excellently with the exact result, while the approximate statistics $n(T)_\mathrm{app}$ overestimates the occupation for temperatures above $T \approx 0.6 \Delta$. For temperatures $T \gg \Delta + W$ (not shown) 
the approximate statistics and the exact curve merge again.

\subsection{Details of the diagrammatic approach}
\label{ssec:details_diagrammatic}

In the previous subsection we  gauged the diagrammatic approach by comparing it to  the exact fermionic results.
In the present subsection, we discuss the results of the diagrammatic expansion for various temperatures and band-widths in more detail. As before we restrict ourselves to the self-consistent solutions.

Even though self-consistency improves the results, we still require that the temperature is not too high, so that higher order processes can be neglected. We stress that the single particle gap $\Delta$ is the most important energy scale, but also the band-width $W$ at zero temperature plays an important role. Especially for the same temperature $T$ and gap $\Delta$ the narrow band limit $W \lessapprox \Delta$ and the wide band limit $W \gtrapprox \Delta$ can differ significantly, because narrow bands generically allow for a much larger fraction of thermal excitations, i.e., a higher density of 
thermally excited hard-core bosons, than wide bands.

\begin{figure}
\centering
\includegraphics[width=1.0\columnwidth]{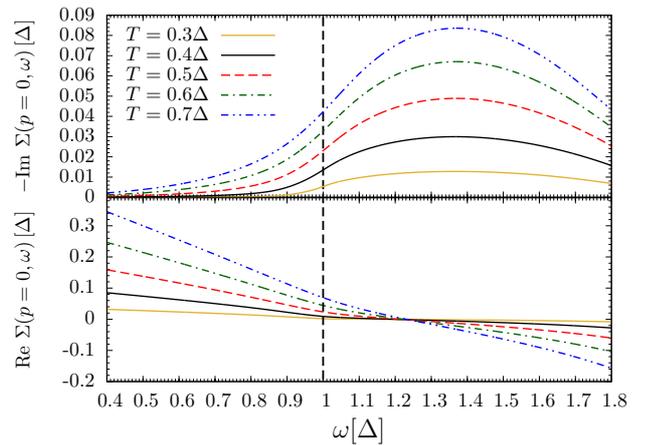}
\caption{{(Color online)} Real and imaginary part of the self energy $\Sigma(p,\omega)$ at various temperatures for the gap mode $p=0$ for $W=0.5\Delta$. The vertical line indicates the energy of the $p=0$ mode at zero temperature. }
\label{fig.ImRe_Sigma_p0_SC_Tdep}
\end{figure}

\begin{figure}
\centering
\includegraphics[width=1.0\columnwidth]{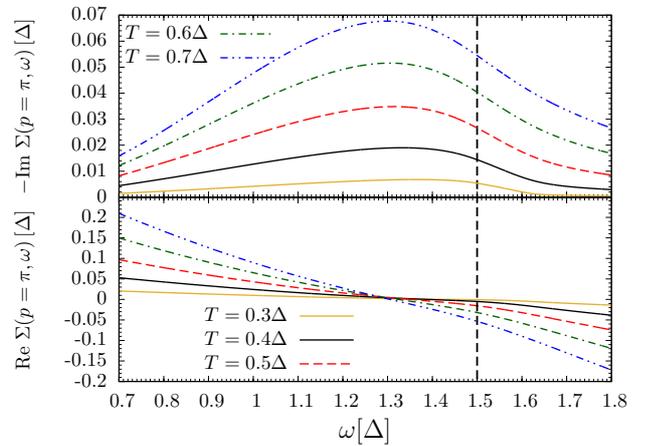}
\caption{{(Color online)} Real and imaginary part of the self energy $\Sigma(p,\omega)$ at various temperatures for the maximum mode $p=\pi$ for $W=0.5\Delta$. The vertical line indicates the energy of the $p=\pi$ mode at zero temperature. }
\label{fig.ImRe_Sigma_p1_SC_Tdep}
\end{figure}

In Figs.\ \ref{fig.ImRe_Sigma_p0_SC_Tdep} and \ref{fig.ImRe_Sigma_p1_SC_Tdep} we investigate the real and the imaginary part of the self energy for the gap mode and for the maximum mode, respectively, in case of the narrow band $W=0.5\Delta$. The imaginary part is dominated by the two-particle continuum convoluted with the single particle Green function, see Eq.\ \eqref{eq.Sigma_convolved}. Upon increasing temperature the imaginary part gains weight. We clearly see for $\omega = \omega(p)$, that the imaginary part is not well approximated by a constant but rather by a linear function, leading to an asymmetric line-shape of the spectral function $A(p, \omega)$. While the gap mode shows a tail towards higher energies due to the positive slope of the imaginary part of the self energy, the maximum mode shows a tail towards lower energies, induced by the negative slope.

The real part of the self energy is dominated by the term $\propto \omega$ in Eq.\ \eqref{eq.sigma_real}, for very high and very low energies. For $p=0$ the real part is positive, indicating a shift of the peak position towards higher energies, while for $p=\pi$ the real part is negative, leading to a shift towards lower energies. With increasing temperatures the effect is amplified due to the increased scattering from thermal excitations.

\subsection{Finite temperature peak broadening}
\label{ssec:broadening}

At zero temperature the single-particle spectral function $A(p, \omega)$ is a $\delta$-function in frequency, signaling stable quasi-particles and providing the dominant contributions to the DSF. Our approach includes this basic property, because the expression for the self energy $\Sigma(p,\omega)$ in Eqs.\ \eqref{eq.sigma_real} and \eqref{eq.sigma_imag} vanishes for $T \rightarrow 0$. Only because our numerical implementation to calculate $A(p, \omega)$ is restricted to a finite frequency and momentum resolution, we require that the temperature is not too low. Otherwise, we are not able to resolve the spectral function. Especially the self-consistent solutions require a finite  broadening as initial input. This inital broadening does not matter at all once the computation is iterated for self-consistency and convergence is indeed reached. 
In our case, using up to $8$ GB of memory to resolve the spectral function, we require that 
$T > 0.15 \Delta$. 

\begin{figure}
\centering
\includegraphics[width=1.0\columnwidth]{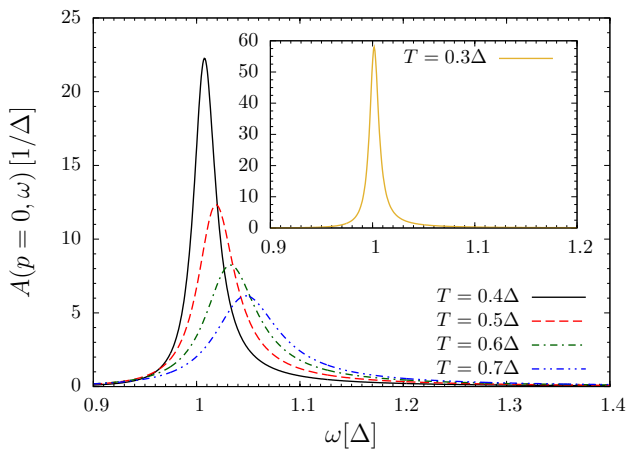}
\caption{{(Color online)} Spectral function $A(p,\omega)$ at various temperatures for the gap mode $p=0$ for $W=0.5\Delta$.}
\label{fig.imG_p0_SC_Tdep}
\end{figure}

First, we consider the temperature dependence of the gap mode at momentum $p=0$ for the narrow band case $W=0.5\Delta$ calculated self-consistently in Fig.\ \ref{fig.imG_p0_SC_Tdep}. For low temperatures the response is primarily of Lorentzian shape and centered at the zero temperature response $\omega = \Delta$. This changes distinctively at higher temperatures. Due to increased scattering with thermal excitations the height of the spectral function decreases. The decrease scales with $\exp(-\beta \Delta)$. The maximum of the response shifts towards higher energies and the peak broadens asymmetrically towards higher energies.

\begin{figure}
\centering
\includegraphics[width=1.0\columnwidth]{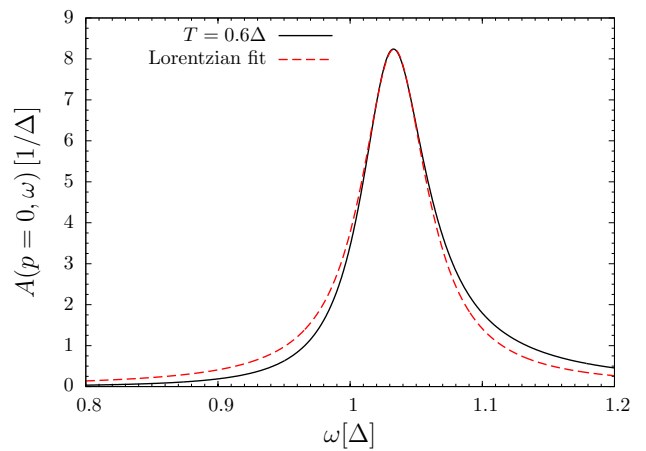}
\caption{{(Color online)} Comparison between Lorentzian line-shape and asymmetric spectral function at $p=0$ for $W=0.5\Delta$.}
\label{fig.imG_p0_SC_Lorentzian}
\end{figure}

The degree of asymmetry is clearly visible in Fig.\ \ref{fig.imG_p0_SC_Lorentzian} where we compare the spectral function to a Lorentzian fit. While a pure Lorentzian shape provides 
evidence for incoherent scattering \cite{Sachdev98}, i.e., an exponential decay in the time domain, asymmetric deviations imply non-trivial scattering. This observation agrees with recent theoretical as well as experimental studies  \cite{ruegg05,mikeska06,essler08a,essler08b,essler09a,essler10a,nafra11,fisch11a,tennant12a,lake12a}.

\begin{figure}
\centering
\includegraphics[width=1.0\columnwidth]{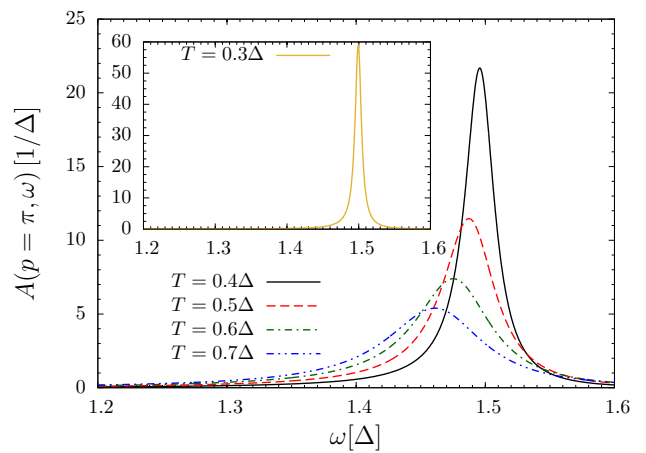}
\caption{{(Color online)} Spectral function $A(p,\omega)$ at various temperatures for the maximum mode $p=\pi$ for $W=0.5\Delta$.}
\label{fig.imG_p1_SC_Tdep}
\end{figure}

Second, we consider the temperature dependence of the maximum mode at momentum $p=\pi$ for the narrow band case $W=0.5\Delta$ calculated self-consistently. This quantity is depicted in Fig.\ \ref{fig.imG_p1_SC_Tdep}. Similar to the gap mode, the spectral function of the maximum mode has a Lorentzian shape and it is centered at the zero temperature position $\omega = \Delta+W$ for low temperatures. The peak broadens with rising temperature and becomes asymmetric. In contrast to the gap mode, the peak position of the maximum mode shifts towards lower energies. Concomitantly, the asymmetric line-shape accumulates weight at lower energies.

\begin{figure}
\centering
\includegraphics[width=1.0\columnwidth]{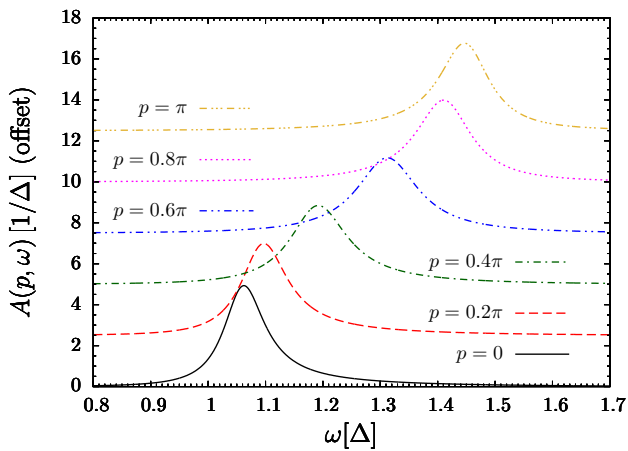}
\caption{{(Color online)} Spectral function for various momenta at fixed temperature $T=0.8\Delta$ for $W=0.5\Delta$. Note the offset of the y-axis.}
\label{fig.imG_T04_SC_pdep}
\end{figure}

An overview plot of the self-consistent spectral function $A(p, \omega)$ at fixed temperature $T=0.8\Delta$ is given in Fig.\ \ref{fig.imG_T04_SC_pdep}. One clearly sees how the asymmetry slowly changes from the gap mode to the maximum mode in dependence of total momentum $p$. In the center of the dispersion, at $p=0.5\pi$, the response remains symmetric and does not change its position with respect to the $T=0$ response.

\subsection{Band narrowing}
\label{ssec:band_narrowing}

\begin{figure}
\centering
\includegraphics[width=1.0\columnwidth]{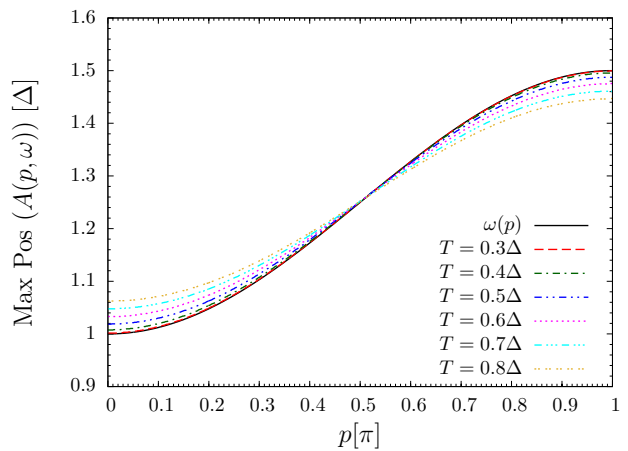}
\caption{{(Color online)} Dispersion determined from the position of the maximum of the spectral function for $W=0.5\Delta$ for various temperatures.}
\label{fig.dispersion_Tdep}
\end{figure}

\begin{figure}
\centering
\includegraphics[width=1.0\columnwidth]{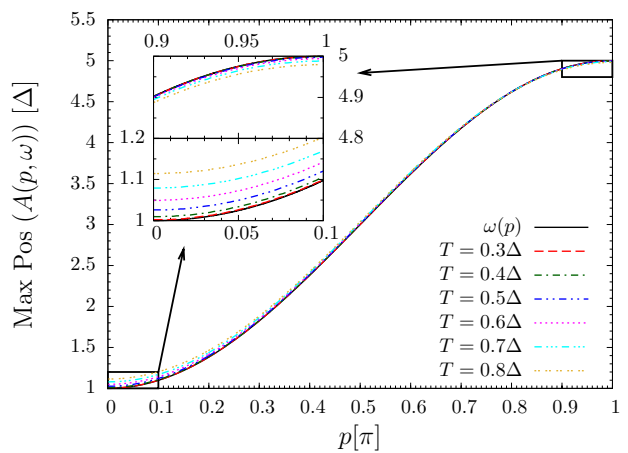}
\caption{{(Color online)} Dispersion determined from the position of the maximum of the spectral function for $W=4\Delta$ for various temperatures. The inset shows a more detailed plot for the gap mode and for the maximum mode.}
\label{fig.dispersion_Tdep_W4}
\end{figure}

The asymmetry and the shift of positions observed in the previous subsection indicates that the total band-width of the system decreases upon rising temperature.
We examine this feature in more detail in Figs.\ \ref{fig.dispersion_Tdep} and \ref{fig.dispersion_Tdep_W4} for the narrow and the wide band case, respectively.
The position of the maximum in the response is plotted as function of the total momentum $p$ for various temperatures $T<\Delta$. While for low temperatures the maximum is located at the dispersion $\omega(p)$, the band is narrowing upon increasing temperature. This effect can be explained by the thermal occupation of an increasing number of sites, blocking the propagation of {an} inserted particle. Consequently, the energy gap is increased and the band-width is decreased. In the literature this is often called the temperature dependence of the gap \cite{exius10} and of the dispersion{,}
 although the physical parameters $\Delta$  and $W$ of the model do not change with temperature in the strict sense. In contrast to the narrow band case $W=0.5\Delta$, the wide band case $W=4\Delta$ indicates that the shift is significantly stronger for the gap mode than for the maximum mode.

\section{Conclusion}

In this paper, we benchmarked a diagrammatic approach to particle-conserving models which  is
capable of dealing with thermal fluctuations in leading order in $\exp(-\beta\Delta)$. 
We stress that particle-conserving effective models are free from quantum fluctuations at zero temperature because
the vacuum is the ground state. Such effective models can be systematically
derived for a large class of microscopic Hamiltonians such as gapped quantum antiferromagnets,
for instance dimerized spin chains \cite{Knetter00,cavad00,schmi03c,tennant12a}, 
spin ladders \cite{schmi05b,normand11}, two-dimensionally coupled spin ladders \cite{uhrig98c,uhrig04a,nafra11,fisch11a}, and three-dimensionally coupled spin dimers \cite{ruegg05,lake12a}.

For the benchmark, we study the spectral properties of an exactly solvable 
gapped one-dimensional hard-core boson model with nearest-neighbor hopping and no
other interaction but the hard-core repulsion. We used diagrammatic perturbation theory to calculate the single-particle self energy. The hard-core constraint was enforced by an infinite on-site interaction. The small control parameter is the density of thermal excitations, which implies that the summation of ladder diagrams is necessary and sufficient to capture the leading order in $\exp(-\beta \Delta)$.  We emphasize that this concept applies equally in any other dimension which
is a particular asset of the approach advocated and tested here.

Our results show how the single-particle $\delta$-peak in the spectral function
broadens with increasing temperature and how band-width and momentum influence the spectral function. We used the mapping to interaction-free Jordan-Wigner fermions to obtain
results which are numerically exact except for finite size and time effects. By this data we
have gauged our diagrammatic approach {and}
evaluated its limits. For low temperature very good agreement is reached. Both methods show that the shift of the peak position is a second order effect $\propto \exp(-2\beta \Delta)$ for the cosine band. In contrast, the width of the peak is a first order effect $\propto \exp(-\beta \Delta)$. 
Thus, it is captured very well by the diagrammatic approach for low enough temperatures. 
Our findings agree with those by Essler and co-workers obtained by direct computations of the leading
contributions to the partition sums \cite{essler08a,essler09a,essler09b}. The focus of the Refs.\ \cite{essler08a,essler09a}
is on the relevant continuum field-theoretical model while Ref.\ \cite{essler09b} studied the strong Ising limit
of a spin chain. The former references deal with an infinitely wide band without lattice while the latter treats a very
narrow band on a lattice. Both studies are specific to one dimension.

With the help of the spectral function we also calculated the thermal occupation,
that is, the total number of thermally excited bosons. The corresponding 
results of the self-consistent 
diagrammatic approach are in excellent agreement with the exact analytic expression available from the
fermionic description,  even for temperatures far above the gap $\Delta$.

We found clear evidence that (i) the line-shapes become asymmetric 
and that (ii) the band width of the overall dispersion narrows upon increasing temperature.
Note that both phenomena seem to be not a specific feature of one dimension but also show up in 
quantum magnets which are coupled in all three space dimensions. One very recent example is given in Ref.\ \cite{lake12a}, where the asymmetric broadening is observed by inelastic neutron scattering in the three dimensional antiferromagnet $\mathrm{Sr}_3\mathrm{Cr}_2\mathrm{O}_8$.

From a theoretical point of view, we expect that the asymmetric line-shapes
as well as the narrowing of the total band width do not depend qualitatively
on dimensionality. The argument in support of this expectation is that
both phenomena are most pronounced in the case of relatively narrow bands
of the order of the gap and at temperatures also of the order of the gap.
In this regime, modes at all momenta are thermally excited and contribute, not
only the modes at the lower band edge. Thus the bulk of the density-of-states
matters. The dimensionality, however, influences mostly the band edges and the
Van Hove singularities at the edges of the density-of-states. 

A different scenario is displayed by systems with a very high or even infinite bandwidth, where the dynamics of the thermal excitations strongly depend on the lower band edge singularity. 
Further calculations for higher dimensional systems are beyond the scope of the
present paper, but they are clearly called for in the near future.

We conclude that the proposed approach is successfully benchmarked against the
exact solution. This finding suggests to generally extend the success of particle-conserving
effective models at zero temperatures to small, but finite temperatures by this diagrammatic
technique. This enlarges the applicability and thus the usefulness of such effective models
considerably. Especially the combination with the CUT approach seems promising. By means of the CUTs, 
the quantum fluctuations of the hard-core bosons at zero temperature are treated in high precision 
by mapping the microscopic models to effective, particle-conserving ones.

We emphasize that the presented diagrammatic  approach does not rely on integrable field theories or other properties specific to one dimension. Thus, various extensions suggest themselves. One may consider more complicated dispersions 
and multi-flavored hard-core bosons in exactly the same way.
Furthermore, one may include other interactions including correlated hopping processes  for the hard-core excitations
at least on a mean-field level. Most importantly, it is conceptually possible to extend the presented approach to higher dimensions and to different models which are currently subject of ongoing
experimental research, see for instance Refs.\ \onlinecite{ruegg05,lake12a}.

We conclude that the diagrammatic expansion is an efficient and universal method to calculate line-shapes at not too high temperatures from the particle-conserving effective models for a wide range of gapped systems with hard-core bosonic excitations. This has been shown in the
present work on the level of a testbed calculation in one dimension. Various extensions can be
tackled next.

\begin{acknowledgments}
We thank T.\ Fischer and F.\ Keim for useful discussions. We acknowledge financial support of the Helmholtz Virtual Institute ``New states of matter and their excitations''. B.F.\ acknowledges the Fakult\"at Physik of the Technische Universit\"at Dortmund for his funding in the program ``Bestenf\"orderung''. 
\end{acknowledgments}

\appendix

\section{Self-consistent calculation}
\label{app.a}

Here, we modifiy the diagrammatic derivations given in Sect.\ \ref{ssect:gen_calcul}
 so that the dressed propagator $G(P)$ is used 
instead of the bare one $G_0(P)$
\begin{align}
G_0(P) \rightarrow G(P) &= \int\limits_{-\infty}^{\infty} \frac{A_p(x)}{i \omega_p - x} \mathrm{d}x .
\end{align}
This changes  the calculation of $M(P)$ in \eqref{eq.M_of_p},
\begin{widetext}
\begin{align}
M(P) &= \frac{1}{N} \sum\limits_{l} \int\limits_{-\infty}^{\infty} \int\limits_{-\infty}^{\infty} \mathrm{d}x \mathrm{d}x'\frac{A_{p+l}(x)A_{-l}(x')}{i\omega_p -(x'+x)} \left( \frac{1}{e^{-\beta x'}-1} - \frac{1}{e^{\beta x}-1} \right),
\end{align}
which leads to a modified expression for $\rho_p(y)$
\begin{align}
\label{eq.rho_p_selfconsistent}
\rho_p(y) &= \frac{-1}{N} \sum\limits_{l} \int\limits_{-\infty}^{\infty} \mathrm{d}x \left( \frac{A_{l}(x)}{e^{\beta x}-1} A_{p-l}(y-x) - A_{l}(x) 
		  	  \frac{A_{p-l}(y-x)}{e^{-\beta(y-x)}-1} \right) .
\end{align}
\end{widetext}
The calculation of $f_p(x)$ in \eqref{eq.calc_f} remains unchanged
so that $\bar{\rho}_p(x)$ now reads
\begin{subequations}
\begin{align}
\label{eq.self_consistent_rho_bar}
\bar{\rho}_p(x) = f_p(x) +  U \left( \frac{\omega_1^2(p)}{\omega_1(p)-\omega_2(p)} \delta(x-\omega_1(p)) \right) .
\end{align}
where
\begin{align}
\omega_1(p) &= - \frac{U \rho_0(p)}{2} + \sqrt{\frac{U^2 \rho_0(p)^2}{4}-U\rho_1(p)}, 
\\
\omega_2(p) &= - \frac{U \rho_0(p)}{2} - \sqrt{\frac{U^2 \rho_0(p)^2}{4}-U\rho_1(p)}. 
\end{align}
\end{subequations}
Due to the self-consistency, the weight function $\rho_0(p)$ now depends on total momentum $p$.
The spectral function for the self energy reads
\begin{widetext}
\begin{align}
\rho_{\Sigma,p}(y) &= \frac{2}{N} \sum\limits_{k} \int\limits_{-\infty}^{\infty} \mathrm{d}x
   \left[ f_{k}(x) \frac{A_{k-p}(x-y)}{e^{\beta \left[x-y\right]} - 1} - \frac{f_{k}(x)}{e^{\beta x} - 1 } A_{k-p}(x-y) \right] ,
\end{align}
\end{widetext}
which is again a two-dimensional convolution
\begin{widetext}
\begin{align}
\label{eq.rho_sigma_selfconsistent}
\rho_{\Sigma,p}(y) &= \frac{2}{N} \sum\limits_{k} \int\limits_{-\infty}^{\infty} \mathrm{d}x
   \left[ f_{k}(x) \frac{A_{-(p-k)}(-(y-x))}{e^{-\beta \left[y-x\right]} - 1} - \frac{f_{k}(x)}{e^{\beta x} - 1 } A_{-(p-k)}(-(y-x)) \right] .
\end{align}
\end{widetext}
The real and imaginary part{s} of the self energy finally read
\begin{subequations}
\begin{widetext}
\begin{align}
\mathrm{Re} \Sigma(p, \omega) &= \frac{2}{N} \sum\limits_k \int\limits_{-\infty}^{\infty} \mathrm{d}x' A_k(x') \left[\frac{\omega}{\rho_0(p+k)}
								 - \frac{\rho_1(p+k)}{\rho_0^2(p+k)} + \frac{x'}{\rho_0(p+k)} \right] \frac{1}{e^{\beta x'}-1} 
					+  \mathcal{P} \int\limits_{-\infty}^{\infty} \frac{\rho_{\Sigma,p}(x)}{\omega-x} \mathrm{d}x ,\\
\mathrm{Im} \Sigma(p, \omega) &= - \pi \rho_{\Sigma,p}(\omega) .
\end{align}
\end{widetext}
\end{subequations}

\section{Computational details}
\label{app.b}

To make the diagrammatic expansion numerically tractable, we discretize all quantities of interest in momentum and frequency space. For momentum space we typically use up to $1024$ points, while up to $65536$ points are used in frequency space to ensure a good resolution. Using smaller numbers does not change the results significantly. The numbers of points are powers of $2$ to make use of fast radix-2-algorithms from the FFTW library \cite{FFTW05}.
The necessary calculations for the diagrammatic expansion of the spectral functions can be divided into four parts.
First the calculation of $\rho_p(x)$ in \eqref{eq.rho_p}. In the non-self-consistent we employ
\begin{subequations}
\begin{align}
 \rho_p(x) &= \frac{-1}{2 \pi} \int\limits_0^{2\pi} \mathrm{d}l \delta(x-[\omega(-l) + \omega(p+l)]) 
 \nonumber \\ &\cdot \left( \frac{1}{e^{\beta \omega(p+l)} - 1} - \frac{1}{e^{- \beta \omega(-l) } - 1} \right) 
 \\
	  &= \frac{-1}{2 \pi} \sum\limits_i \frac{1}{\left| \omega'(-l_i)+\omega'(p+l_i) \right|} 
	  \nonumber\\ 
	  &\cdot \left( \frac{1}{e^{\beta \omega(p+l_i)} - 1} - \frac{1}{e^{- \beta \omega(-l_i) } - 1} \right)
\end{align}
\end{subequations}
where $\omega'(l)$ is the derivative of the dispersion with respect to $l$. The values $l_i$ are the roots of the function $g(l) = x-[\omega(-l) + \omega(p+l)]$. They are calculated numerically using a one-dimensional root finding algorithm. In the self-consistent case the calculation can be carried out as given in \eqref{eq.rho_p_selfconsistent} because it represents a two-dimensional convolution in the variables $l$ and $x$. It can be calculated using conventional, fast convolution algorithms, based on fast Fourier transforms (FFTs).

The second step is the calculation of $f_p(x)$ in \eqref{eq.calc_f}. Here the main intricacy is the calculation of the principal value. This can be carried out efficiently using the tricks proposed by Liu and Kosloff \cite{LiuKosloff81}.

The third step is the calculation of $\rho_{\Sigma,p} (\omega)$ in \eqref{eq.rho_sigma}. In the non-self-consistent case this can be done using standard one-dimensional integration algorithms, while the self-consistent case can again be mapped to a two-dimensional convolution, see \eqref{eq.rho_sigma_selfconsistent}.

Finally, some remarks on the calculation of the real and imaginary part{s} of the self energy. While the imaginary part is again trivial, the real part requires another principal value integral and three one- or two-dimensional integrals. 

It turns out that the calculations based on two-dimensional convolutions are much faster than those in the non-self-consistent case. This can be traced back to the fact that a single two-dimensional convolution is sufficient to calculate quantities such as $\rho_p(x)$, while the same quantity in the non-self-consistent case requires the computation of a full set of roots for each pair of values $x$ and $p$ we are interested in.

\bibliographystyle{apsrev}

\begin{thebibliography}{52}
\expandafter\ifx\csname natexlab\endcsname\relax\def\natexlab#1{#1}\fi
\expandafter\ifx\csname bibnamefont\endcsname\relax
  \def\bibnamefont#1{#1}\fi
\expandafter\ifx\csname bibfnamefont\endcsname\relax
  \def\bibfnamefont#1{#1}\fi
\expandafter\ifx\csname citenamefont\endcsname\relax
  \def\citenamefont#1{#1}\fi
\expandafter\ifx\csname url\endcsname\relax
  \def\url#1{\texttt{#1}}\fi
\expandafter\ifx\csname urlprefix\endcsname\relax\def\urlprefix{URL }\fi
\providecommand{\bibinfo}[2]{#2}
\providecommand{\eprint}[2][]{\url{#2}}

\bibitem[{\citenamefont{S\'olyom}(1979)}]{solyo79}
\bibinfo{author}{\bibfnamefont{J.}~\bibnamefont{S\'olyom}},
  \bibinfo{journal}{Adv. Phys.} \textbf{\bibinfo{volume}{28}},
  \bibinfo{pages}{201} (\bibinfo{year}{1979}).

\bibitem[{\citenamefont{Metzner et~al.}(2012)\citenamefont{Metzner, Salmhofer,
  Honerkamp, Meden, and Sch\"onhammer}}]{metzn12}
\bibinfo{author}{\bibfnamefont{W.}~\bibnamefont{Metzner}},
  \bibinfo{author}{\bibfnamefont{M.}~\bibnamefont{Salmhofer}},
  \bibinfo{author}{\bibfnamefont{C.}~\bibnamefont{Honerkamp}},
  \bibinfo{author}{\bibfnamefont{V.}~\bibnamefont{Meden}}, \bibnamefont{and}
  \bibinfo{author}{\bibfnamefont{K.}~\bibnamefont{Sch\"onhammer}},
  \bibinfo{journal}{Rev. Mod. Phys.} \textbf{\bibinfo{volume}{84}},
  \bibinfo{pages}{299} (\bibinfo{year}{2012}).

\bibitem[{\citenamefont{Knetter et~al.}(2003)\citenamefont{Knetter, Schmidt,
  and Uhrig}}]{knett03a}
\bibinfo{author}{\bibfnamefont{C.}~\bibnamefont{Knetter}},
  \bibinfo{author}{\bibfnamefont{K.~P.} \bibnamefont{Schmidt}},
  \bibnamefont{and} \bibinfo{author}{\bibfnamefont{G.~S.} \bibnamefont{Uhrig}},
  \bibinfo{journal}{J. Phys. A: Math. Gen.} \textbf{\bibinfo{volume}{36}},
  \bibinfo{pages}{7889} (\bibinfo{year}{2003}).

\bibitem[{\citenamefont{Haegeman et~al.}(2013)\citenamefont{Haegeman,
  Michalakis, Nachtergaele, Osborne, Schuch, and Verstraete}}]{haege13a}
\bibinfo{author}{\bibfnamefont{J.}~\bibnamefont{Haegeman}},
  \bibinfo{author}{\bibfnamefont{S.}~\bibnamefont{Michalakis}},
  \bibinfo{author}{\bibfnamefont{B.}~\bibnamefont{Nachtergaele}},
  \bibinfo{author}{\bibfnamefont{T.~J.} \bibnamefont{Osborne}},
  \bibinfo{author}{\bibfnamefont{N.}~\bibnamefont{Schuch}}, \bibnamefont{and}
  \bibinfo{author}{\bibfnamefont{F.}~\bibnamefont{Verstraete}},
  \bibinfo{journal}{Phys. Rev. Lett.} \textbf{\bibinfo{volume}{111}},
  \bibinfo{pages}{080401} (\bibinfo{year}{2013}).

\bibitem[{\citenamefont{{C. Knetter and G.S. Uhrig}}(2000)}]{Knetter00}
\bibinfo{author}{\bibnamefont{{C. Knetter and G.S. Uhrig}}},
  \bibinfo{journal}{Eur. Phys. J. B} \textbf{\bibinfo{volume}{13}},
  \bibinfo{pages}{209} (\bibinfo{year}{2000}).

\bibitem[{\citenamefont{Cavadini et~al.}(2000)\citenamefont{Cavadini, R\"uegg,
  Henggeler, Furrer, G\"udel, Kr\"amer, and Mutka}}]{cavad00}
\bibinfo{author}{\bibfnamefont{N.}~\bibnamefont{Cavadini}},
  \bibinfo{author}{\bibfnamefont{C.}~\bibnamefont{R\"uegg}},
  \bibinfo{author}{\bibfnamefont{W.}~\bibnamefont{Henggeler}},
  \bibinfo{author}{\bibfnamefont{A.}~\bibnamefont{Furrer}},
  \bibinfo{author}{\bibfnamefont{H.-U.} \bibnamefont{G\"udel}},
  \bibinfo{author}{\bibfnamefont{K.}~\bibnamefont{Kr\"amer}}, \bibnamefont{and}
  \bibinfo{author}{\bibfnamefont{H.}~\bibnamefont{Mutka}},
  \bibinfo{journal}{Eur. Phys. J. B} \textbf{\bibinfo{volume}{18}},
  \bibinfo{pages}{565} (\bibinfo{year}{2000}).

\bibitem[{\citenamefont{Tennant et~al.}(2012)\citenamefont{Tennant, Lake,
  James, Essler, Notbohm, Mikeska, Fielden, K\"ogerler, Canfield, and
  Telling}}]{tennant12a}
\bibinfo{author}{\bibfnamefont{D.~A.} \bibnamefont{Tennant}},
  \bibinfo{author}{\bibfnamefont{B.}~\bibnamefont{Lake}},
  \bibinfo{author}{\bibfnamefont{A.~J.~A.} \bibnamefont{James}},
  \bibinfo{author}{\bibfnamefont{F.~H.~L.} \bibnamefont{Essler}},
  \bibinfo{author}{\bibfnamefont{S.}~\bibnamefont{Notbohm}},
  \bibinfo{author}{\bibfnamefont{H.-J.} \bibnamefont{Mikeska}},
  \bibinfo{author}{\bibfnamefont{J.}~\bibnamefont{Fielden}},
  \bibinfo{author}{\bibfnamefont{P.}~\bibnamefont{K\"ogerler}},
  \bibinfo{author}{\bibfnamefont{P.~C.} \bibnamefont{Canfield}},
  \bibnamefont{and} \bibinfo{author}{\bibfnamefont{M.~T.~F.}
  \bibnamefont{Telling}}, \bibinfo{journal}{Phys. Rev. B}
  \textbf{\bibinfo{volume}{85}}, \bibinfo{pages}{014402}
  (\bibinfo{year}{2012}).

\bibitem[{\citenamefont{Schmidt and Uhrig}(2005)}]{schmi05b}
\bibinfo{author}{\bibfnamefont{K.~P.} \bibnamefont{Schmidt}} \bibnamefont{and}
  \bibinfo{author}{\bibfnamefont{G.~S.} \bibnamefont{Uhrig}},
  \bibinfo{journal}{Mod. Phys. Lett. B} \textbf{\bibinfo{volume}{19}},
  \bibinfo{pages}{1179} (\bibinfo{year}{2005}).

\bibitem[{\citenamefont{Normand and R\"uegg}(2011)}]{normand11}
\bibinfo{author}{\bibfnamefont{B.}~\bibnamefont{Normand}} \bibnamefont{and}
  \bibinfo{author}{\bibfnamefont{C.}~\bibnamefont{R\"uegg}},
  \bibinfo{journal}{Phys. Rev. B} \textbf{\bibinfo{volume}{83}},
  \bibinfo{pages}{054415} (\bibinfo{year}{2011}).

\bibitem[{\citenamefont{Uhrig and Normand}(1998)}]{uhrig98c}
\bibinfo{author}{\bibfnamefont{G.~S.} \bibnamefont{Uhrig}} \bibnamefont{and}
  \bibinfo{author}{\bibfnamefont{B.}~\bibnamefont{Normand}},
  \bibinfo{journal}{Phys. Rev. B} \textbf{\bibinfo{volume}{58}},
  \bibinfo{pages}{14705(R)} (\bibinfo{year}{1998}).

\bibitem[{\citenamefont{Uhrig et~al.}(2004)\citenamefont{Uhrig, Schmidt, and
  Gr\"uninger}}]{uhrig04a}
\bibinfo{author}{\bibfnamefont{G.~S.} \bibnamefont{Uhrig}},
  \bibinfo{author}{\bibfnamefont{K.~P.} \bibnamefont{Schmidt}},
  \bibnamefont{and}
  \bibinfo{author}{\bibfnamefont{M.}~\bibnamefont{Gr\"uninger}},
  \bibinfo{journal}{Phys. Rev. Lett.} \textbf{\bibinfo{volume}{93}},
  \bibinfo{pages}{267003} (\bibinfo{year}{2004}).

\bibitem[{\citenamefont{N\'afr\'adi et~al.}(2011)\citenamefont{N\'afr\'adi,
  Keller, Manaka, Zheludev, and Keimer}}]{nafra11}
\bibinfo{author}{\bibfnamefont{B.}~\bibnamefont{N\'afr\'adi}},
  \bibinfo{author}{\bibfnamefont{T.}~\bibnamefont{Keller}},
  \bibinfo{author}{\bibfnamefont{H.}~\bibnamefont{Manaka}},
  \bibinfo{author}{\bibfnamefont{A.}~\bibnamefont{Zheludev}}, \bibnamefont{and}
  \bibinfo{author}{\bibfnamefont{B.}~\bibnamefont{Keimer}},
  \bibinfo{journal}{Phys. Rev. Lett.} \textbf{\bibinfo{volume}{106}},
  \bibinfo{pages}{177202} (\bibinfo{year}{2011}).

\bibitem[{\citenamefont{Fischer et~al.}(2011)\citenamefont{Fischer, Duffe, and
  Uhrig}}]{fisch11a}
\bibinfo{author}{\bibfnamefont{T.}~\bibnamefont{Fischer}},
  \bibinfo{author}{\bibfnamefont{S.}~\bibnamefont{Duffe}}, \bibnamefont{and}
  \bibinfo{author}{\bibfnamefont{G.~S.} \bibnamefont{Uhrig}},
  \bibinfo{journal}{Europhys. Lett.} \textbf{\bibinfo{volume}{96}},
  \bibinfo{pages}{47001} (\bibinfo{year}{2011}).

\bibitem[{\citenamefont{R\"uegg et~al.}(2005)\citenamefont{R\"uegg, Normand,
  Matsumoto, Niedermayer, Furrer, Kr\"amer, G\"udel, Bourges, Sidis, and
  Mutka}}]{ruegg05}
\bibinfo{author}{\bibfnamefont{C.}~\bibnamefont{R\"uegg}},
  \bibinfo{author}{\bibfnamefont{B.}~\bibnamefont{Normand}},
  \bibinfo{author}{\bibfnamefont{M.}~\bibnamefont{Matsumoto}},
  \bibinfo{author}{\bibfnamefont{C.}~\bibnamefont{Niedermayer}},
  \bibinfo{author}{\bibfnamefont{A.}~\bibnamefont{Furrer}},
  \bibinfo{author}{\bibfnamefont{K.~W.} \bibnamefont{Kr\"amer}},
  \bibinfo{author}{\bibfnamefont{H.-U.} \bibnamefont{G\"udel}},
  \bibinfo{author}{\bibfnamefont{P.}~\bibnamefont{Bourges}},
  \bibinfo{author}{\bibfnamefont{Y.}~\bibnamefont{Sidis}}, \bibnamefont{and}
  \bibinfo{author}{\bibfnamefont{H.}~\bibnamefont{Mutka}},
  \bibinfo{journal}{Phys. Rev. Lett.} \textbf{\bibinfo{volume}{95}},
  \bibinfo{pages}{267201} (\bibinfo{year}{2005}).

\bibitem[{\citenamefont{Quintero-Castro
  et~al.}(2012)\citenamefont{Quintero-Castro, Lake, Islam, Wheeler, Balz,
  M\aa{}nsson, Rule, Gvasaliya, and Zheludev}}]{lake12a}
\bibinfo{author}{\bibfnamefont{D.~L.} \bibnamefont{Quintero-Castro}},
  \bibinfo{author}{\bibfnamefont{B.}~\bibnamefont{Lake}},
  \bibinfo{author}{\bibfnamefont{A.~T. M.~N.} \bibnamefont{Islam}},
  \bibinfo{author}{\bibfnamefont{E.~M.} \bibnamefont{Wheeler}},
  \bibinfo{author}{\bibfnamefont{C.}~\bibnamefont{Balz}},
  \bibinfo{author}{\bibfnamefont{M.}~\bibnamefont{M\aa{}nsson}},
  \bibinfo{author}{\bibfnamefont{K.~C.} \bibnamefont{Rule}},
  \bibinfo{author}{\bibfnamefont{S.}~\bibnamefont{Gvasaliya}},
  \bibnamefont{and} \bibinfo{author}{\bibfnamefont{A.}~\bibnamefont{Zheludev}},
  \bibinfo{journal}{Phys. Rev. Lett.} \textbf{\bibinfo{volume}{109}},
  \bibinfo{pages}{127206} (\bibinfo{year}{2012}).

\bibitem[{\citenamefont{Schmidt and Uhrig}(2003)}]{schmi03c}
\bibinfo{author}{\bibfnamefont{K.~P.} \bibnamefont{Schmidt}} \bibnamefont{and}
  \bibinfo{author}{\bibfnamefont{G.~S.} \bibnamefont{Uhrig}},
  \bibinfo{journal}{Phys. Rev. Lett.} \textbf{\bibinfo{volume}{90}},
  \bibinfo{pages}{227204} (\bibinfo{year}{2003}).

\bibitem[{\citenamefont{Fauseweh and Uhrig}(2013)}]{fause13a}
\bibinfo{author}{\bibfnamefont{B.}~\bibnamefont{Fauseweh}} \bibnamefont{and}
  \bibinfo{author}{\bibfnamefont{G.~S.} \bibnamefont{Uhrig}},
  \bibinfo{journal}{Phys. Rev. B} \textbf{\bibinfo{volume}{87}},
  \bibinfo{pages}{184406} (\bibinfo{year}{2013}).

\bibitem[{\citenamefont{Bloch et~al.}(2008)\citenamefont{Bloch, Dalibard, and
  Zwerger}}]{bloch08}
\bibinfo{author}{\bibfnamefont{I.}~\bibnamefont{Bloch}},
  \bibinfo{author}{\bibfnamefont{J.}~\bibnamefont{Dalibard}}, \bibnamefont{and}
  \bibinfo{author}{\bibfnamefont{W.}~\bibnamefont{Zwerger}},
  \bibinfo{journal}{Rev. Mod. Phys.} \textbf{\bibinfo{volume}{80}},
  \bibinfo{pages}{885} (\bibinfo{year}{2008}).

\bibitem[{\citenamefont{Schmidt et~al.}(2004)\citenamefont{Schmidt, Knetter,
  and Uhrig}}]{schmi04a}
\bibinfo{author}{\bibfnamefont{K.~P.} \bibnamefont{Schmidt}},
  \bibinfo{author}{\bibfnamefont{C.}~\bibnamefont{Knetter}}, \bibnamefont{and}
  \bibinfo{author}{\bibfnamefont{G.~S.} \bibnamefont{Uhrig}},
  \bibinfo{journal}{Phys. Rev. B} \textbf{\bibinfo{volume}{69}},
  \bibinfo{pages}{104417} (\bibinfo{year}{2004}).

\bibitem[{\citenamefont{Mikeska and Luckmann}(2006)}]{mikeska06}
\bibinfo{author}{\bibfnamefont{H.~J.} \bibnamefont{Mikeska}} \bibnamefont{and}
  \bibinfo{author}{\bibfnamefont{C.}~\bibnamefont{Luckmann}},
  \bibinfo{journal}{Phys. Rev. B} \textbf{\bibinfo{volume}{73}},
  \bibinfo{pages}{184426} (\bibinfo{year}{2006}).

\bibitem[{\citenamefont{Essler and Konik}(2008)}]{essler08a}
\bibinfo{author}{\bibfnamefont{F.~H.~L.} \bibnamefont{Essler}}
  \bibnamefont{and} \bibinfo{author}{\bibfnamefont{R.~M.} \bibnamefont{Konik}},
  \bibinfo{journal}{Phys. Rev. B} \textbf{\bibinfo{volume}{78}},
  \bibinfo{pages}{100403} (\bibinfo{year}{2008}).

\bibitem[{\citenamefont{James et~al.}(2008)\citenamefont{James, Essler, and
  Konik}}]{essler08b}
\bibinfo{author}{\bibfnamefont{A.~J.~A.} \bibnamefont{James}},
  \bibinfo{author}{\bibfnamefont{F.~H.~L.} \bibnamefont{Essler}},
  \bibnamefont{and} \bibinfo{author}{\bibfnamefont{R.~M.} \bibnamefont{Konik}},
  \bibinfo{journal}{Phys. Rev. B} \textbf{\bibinfo{volume}{78}},
  \bibinfo{pages}{094411} (\bibinfo{year}{2008}).

\bibitem[{\citenamefont{Essler and Konik}(2009)}]{essler09a}
\bibinfo{author}{\bibfnamefont{F.~H.~L.} \bibnamefont{Essler}}
  \bibnamefont{and} \bibinfo{author}{\bibfnamefont{R.~M.} \bibnamefont{Konik}},
  \bibinfo{journal}{J. Stat. Mech.: Theor. Exp.} p. \bibinfo{pages}{P09018}
  (\bibinfo{year}{2009}).

\bibitem[{\citenamefont{Goetze et~al.}(2010)\citenamefont{Goetze,
  Karahasanovic, and Essler}}]{essler10a}
\bibinfo{author}{\bibfnamefont{W.~D.} \bibnamefont{Goetze}},
  \bibinfo{author}{\bibfnamefont{U.}~\bibnamefont{Karahasanovic}},
  \bibnamefont{and} \bibinfo{author}{\bibfnamefont{F.~H.~L.}
  \bibnamefont{Essler}}, \bibinfo{journal}{Phys. Rev. B}
  \textbf{\bibinfo{volume}{82}}, \bibinfo{pages}{104417}
  (\bibinfo{year}{2010}).

\bibitem[{\citenamefont{Wegner}(1994)}]{Wegner94}
\bibinfo{author}{\bibfnamefont{F.}~\bibnamefont{Wegner}},
  \bibinfo{journal}{Ann. Physik} \textbf{\bibinfo{volume}{506}},
  \bibinfo{pages}{77} (\bibinfo{year}{1994}).

\bibitem[{\citenamefont{{S. D. G\l{}azek and K. G.
  Wilson}}(1993)}]{GlazekWilson93}
\bibinfo{author}{\bibnamefont{{S. D. G\l{}azek and K. G. Wilson}}},
  \bibinfo{journal}{Phys. Rev. D} \textbf{\bibinfo{volume}{48}},
  \bibinfo{pages}{5863} (\bibinfo{year}{1993}).

\bibitem[{\citenamefont{{S. D. G\l{}azek and K. G.
  Wilson}}(1994)}]{GlazekWilson94}
\bibinfo{author}{\bibnamefont{{S. D. G\l{}azek and K. G. Wilson}}},
  \bibinfo{journal}{Phys. Rev. D} \textbf{\bibinfo{volume}{49}},
  \bibinfo{pages}{4214} (\bibinfo{year}{1994}).

\bibitem[{\citenamefont{Fischer et~al.}(2010)\citenamefont{Fischer, Duffe, and
  Uhrig}}]{Fischer2010}
\bibinfo{author}{\bibfnamefont{T.}~\bibnamefont{Fischer}},
  \bibinfo{author}{\bibfnamefont{S.}~\bibnamefont{Duffe}}, \bibnamefont{and}
  \bibinfo{author}{\bibfnamefont{G.~S.} \bibnamefont{Uhrig}},
  \bibinfo{journal}{New J. Phys.} \textbf{\bibinfo{volume}{12}},
  \bibinfo{pages}{033048} (\bibinfo{year}{2010}).

\bibitem[{\citenamefont{{H. Krull, N. A. Drescher and G. S.
  Uhrig}}(2012)}]{Krull12}
\bibinfo{author}{\bibnamefont{{H. Krull, N. A. Drescher and G. S. Uhrig}}},
  \bibinfo{journal}{Phys. Rev. B} \textbf{\bibinfo{volume}{86}},
  \bibinfo{pages}{125113} (\bibinfo{year}{2012}).

\bibitem[{\citenamefont{{P. Jordan and E. Wigner}}(1928)}]{JordanWigner28}
\bibinfo{author}{\bibnamefont{{P. Jordan and E. Wigner}}}, \bibinfo{journal}{Z.
  Phys.} \textbf{\bibinfo{volume}{47}}, \bibinfo{pages}{631}
  (\bibinfo{year}{1928}).

\bibitem[{\citenamefont{Parkinson and
  Farnell}(2010)}]{ParkinsonFarnell:quantumspinsystems}
\bibinfo{author}{\bibfnamefont{J.~B.} \bibnamefont{Parkinson}}
  \bibnamefont{and} \bibinfo{author}{\bibfnamefont{D.~J.~J.}
  \bibnamefont{Farnell}}, \emph{\bibinfo{title}{{An Introduction to Quantum
  Spin Systems}}} (\bibinfo{publisher}{Springer}, \bibinfo{address}{New York,
  N.Y.}, \bibinfo{year}{2010}), ISBN \bibinfo{isbn}{978-3-642-13289-6}.

\bibitem[{\citenamefont{Fetter and Walecka}(2003)}]{fetterwalecka}
\bibinfo{author}{\bibfnamefont{A.~L.} \bibnamefont{Fetter}} \bibnamefont{and}
  \bibinfo{author}{\bibfnamefont{J.~D.} \bibnamefont{Walecka}},
  \emph{\bibinfo{title}{{Quantum Theory of Many-Particle Systems}}}
  (\bibinfo{publisher}{Dover Publications}, \bibinfo{year}{2003}), ISBN
  \bibinfo{isbn}{0486428273}.

\bibitem[{\citenamefont{Kotov et~al.}(1998)\citenamefont{Kotov, Sushkov,
  Weihong, and Oitmaa}}]{kotov98a}
\bibinfo{author}{\bibfnamefont{V.~N.} \bibnamefont{Kotov}},
  \bibinfo{author}{\bibfnamefont{O.}~\bibnamefont{Sushkov}},
  \bibinfo{author}{\bibfnamefont{Z.}~\bibnamefont{Weihong}}, \bibnamefont{and}
  \bibinfo{author}{\bibfnamefont{J.}~\bibnamefont{Oitmaa}},
  \bibinfo{journal}{Phys. Rev. Lett.} \textbf{\bibinfo{volume}{80}},
  \bibinfo{pages}{5790} (\bibinfo{year}{1998}).

\bibitem[{\citenamefont{Abrikosov et~al.}(1975)\citenamefont{Abrikosov, Gorkov,
  and Dzyaloshinski}}]{Abrikosov:QFT}
\bibinfo{author}{\bibfnamefont{A.}~\bibnamefont{Abrikosov}},
  \bibinfo{author}{\bibfnamefont{L.}~\bibnamefont{Gorkov}}, \bibnamefont{and}
  \bibinfo{author}{\bibfnamefont{I.}~\bibnamefont{Dzyaloshinski}},
  \emph{\bibinfo{title}{{Methods of quantum field theory in statistical
  physics}}} (\bibinfo{publisher}{Dover}, \bibinfo{address}{New York, N.Y.},
  \bibinfo{year}{1975}), ISBN \bibinfo{isbn}{978-0486632285}.

\bibitem[{\citenamefont{Shevchenko et~al.}(2000)\citenamefont{Shevchenko,
  Sandvik, and Sushkov}}]{sushkov00}
\bibinfo{author}{\bibfnamefont{P.~V.} \bibnamefont{Shevchenko}},
  \bibinfo{author}{\bibfnamefont{A.~W.} \bibnamefont{Sandvik}},
  \bibnamefont{and} \bibinfo{author}{\bibfnamefont{O.~P.}
  \bibnamefont{Sushkov}}, \bibinfo{journal}{Phys. Rev. B}
  \textbf{\bibinfo{volume}{61}}, \bibinfo{pages}{3475} (\bibinfo{year}{2000}).

\bibitem[{\citenamefont{Baym and Kadanoff}(1961)}]{baym61}
\bibinfo{author}{\bibfnamefont{G.}~\bibnamefont{Baym}} \bibnamefont{and}
  \bibinfo{author}{\bibfnamefont{L.~P.} \bibnamefont{Kadanoff}},
  \bibinfo{journal}{Phys. Rev.} \textbf{\bibinfo{volume}{124}},
  \bibinfo{pages}{287} (\bibinfo{year}{1961}).

\bibitem[{\citenamefont{Baym}(1962)}]{baym62}
\bibinfo{author}{\bibfnamefont{G.}~\bibnamefont{Baym}}, \bibinfo{journal}{Phys.
  Rev.} \textbf{\bibinfo{volume}{127}}, \bibinfo{pages}{1391}
  (\bibinfo{year}{1962}).

\bibitem[{\citenamefont{Lieb et~al.}(1961)\citenamefont{Lieb, Schultz, and
  Mattis}}]{LSM61}
\bibinfo{author}{\bibfnamefont{E.}~\bibnamefont{Lieb}},
  \bibinfo{author}{\bibfnamefont{T.}~\bibnamefont{Schultz}}, \bibnamefont{and}
  \bibinfo{author}{\bibfnamefont{D.}~\bibnamefont{Mattis}},
  \bibinfo{journal}{Ann. of Phys.} \textbf{\bibinfo{volume}{16}},
  \bibinfo{pages}{407} (\bibinfo{year}{1961}).

\bibitem[{\citenamefont{Katsura}(1962)}]{Kat62}
\bibinfo{author}{\bibfnamefont{S.}~\bibnamefont{Katsura}},
  \bibinfo{journal}{Phys. Rev.} \textbf{\bibinfo{volume}{127}},
  \bibinfo{pages}{1508} (\bibinfo{year}{1962}).

\bibitem[{\citenamefont{Gaudin}(1960)}]{Gau60}
\bibinfo{author}{\bibfnamefont{M.}~\bibnamefont{Gaudin}},
  \bibinfo{journal}{Nucl. Phys.} \textbf{\bibinfo{volume}{15}},
  \bibinfo{pages}{89} (\bibinfo{year}{1960}).

\bibitem[{\citenamefont{Green and Hurst}(1964)}]{GH64}
\bibinfo{author}{\bibfnamefont{H.~S.} \bibnamefont{Green}} \bibnamefont{and}
  \bibinfo{author}{\bibfnamefont{C.~A.} \bibnamefont{Hurst}},
  \emph{\bibinfo{title}{Order-Disorder Phenomena}}
  (\bibinfo{publisher}{Wiley-Interscience}, \bibinfo{address}{London},
  \bibinfo{year}{1964}).

\bibitem[{\citenamefont{Derzhko and Krokhmalskii}(1998)}]{DK98}
\bibinfo{author}{\bibfnamefont{O.}~\bibnamefont{Derzhko}} \bibnamefont{and}
  \bibinfo{author}{\bibfnamefont{T.}~\bibnamefont{Krokhmalskii}},
  \bibinfo{journal}{phys. stat. sol. (b)} \textbf{\bibinfo{volume}{208}},
  \bibinfo{pages}{221} (\bibinfo{year}{1998}).

\bibitem[{\citenamefont{Jia and Chakravarty}(2006)}]{JC06}
\bibinfo{author}{\bibfnamefont{X.}~\bibnamefont{Jia}} \bibnamefont{and}
  \bibinfo{author}{\bibfnamefont{S.}~\bibnamefont{Chakravarty}},
  \bibinfo{journal}{Phys. Rev. B} \textbf{\bibinfo{volume}{74}},
  \bibinfo{pages}{172414} (\bibinfo{year}{2006}).

\bibitem[{\citenamefont{Cruz and Gon\c{c}alves}(1981)}]{CG81}
\bibinfo{author}{\bibfnamefont{H.~B.} \bibnamefont{Cruz}} \bibnamefont{and}
  \bibinfo{author}{\bibfnamefont{L.~L.} \bibnamefont{Gon\c{c}alves}},
  \bibinfo{journal}{J. Phys. C} \textbf{\bibinfo{volume}{14}},
  \bibinfo{pages}{2785} (\bibinfo{year}{1981}).

\bibitem[{\citenamefont{Derzhko et~al.}(2000)\citenamefont{Derzhko,
  Krokhmalskii, and Stolze}}]{my38}
\bibinfo{author}{\bibfnamefont{O.}~\bibnamefont{Derzhko}},
  \bibinfo{author}{\bibfnamefont{T.}~\bibnamefont{Krokhmalskii}},
  \bibnamefont{and} \bibinfo{author}{\bibfnamefont{J.}~\bibnamefont{Stolze}},
  \bibinfo{journal}{J. Phys. A: Math. Gen.} \textbf{\bibinfo{volume}{33}},
  \bibinfo{pages}{3063} (\bibinfo{year}{2000}).

\bibitem[{\citenamefont{Derzhko et~al.}(2002)\citenamefont{Derzhko,
  Krokhmalskii, and Stolze}}]{my40}
\bibinfo{author}{\bibfnamefont{O.}~\bibnamefont{Derzhko}},
  \bibinfo{author}{\bibfnamefont{T.}~\bibnamefont{Krokhmalskii}},
  \bibnamefont{and} \bibinfo{author}{\bibfnamefont{J.}~\bibnamefont{Stolze}},
  \bibinfo{journal}{J. Phys. A: Math. Gen.} \textbf{\bibinfo{volume}{35}},
  \bibinfo{pages}{3573} (\bibinfo{year}{2002}).

\bibitem[{\citenamefont{Troyer et~al.}(1994)\citenamefont{Troyer, Tsunetsugu,
  and W\"urtz}}]{troyer94}
\bibinfo{author}{\bibfnamefont{M.}~\bibnamefont{Troyer}},
  \bibinfo{author}{\bibfnamefont{H.}~\bibnamefont{Tsunetsugu}},
  \bibnamefont{and} \bibinfo{author}{\bibfnamefont{D.}~\bibnamefont{W\"urtz}},
  \bibinfo{journal}{Phys. Rev. B} \textbf{\bibinfo{volume}{50}},
  \bibinfo{pages}{13515} (\bibinfo{year}{1994}).

\bibitem[{\citenamefont{Exius et~al.}(2010)\citenamefont{Exius, Schmidt, Lake,
  Tennant, and Uhrig}}]{exius10}
\bibinfo{author}{\bibfnamefont{I.}~\bibnamefont{Exius}},
  \bibinfo{author}{\bibfnamefont{K.~P.} \bibnamefont{Schmidt}},
  \bibinfo{author}{\bibfnamefont{B.}~\bibnamefont{Lake}},
  \bibinfo{author}{\bibfnamefont{D.~A.} \bibnamefont{Tennant}},
  \bibnamefont{and} \bibinfo{author}{\bibfnamefont{G.~S.} \bibnamefont{Uhrig}},
  \bibinfo{journal}{Phys. Rev. B} \textbf{\bibinfo{volume}{82}},
  \bibinfo{pages}{214410} (\bibinfo{year}{2010}).

\bibitem[{\citenamefont{Damle and Sachdev}(1998)}]{Sachdev98}
\bibinfo{author}{\bibfnamefont{K.}~\bibnamefont{Damle}} \bibnamefont{and}
  \bibinfo{author}{\bibfnamefont{S.}~\bibnamefont{Sachdev}},
  \bibinfo{journal}{Phys. Rev. B} \textbf{\bibinfo{volume}{57}},
  \bibinfo{pages}{8307} (\bibinfo{year}{1998}).

\bibitem[{\citenamefont{James et~al.}(2009)\citenamefont{James, Goetze, and
  Essler}}]{essler09b}
\bibinfo{author}{\bibfnamefont{A.~J.~A.} \bibnamefont{James}},
  \bibinfo{author}{\bibfnamefont{W.~D.} \bibnamefont{Goetze}},
  \bibnamefont{and} \bibinfo{author}{\bibfnamefont{F.~H.~L.}
  \bibnamefont{Essler}}, \bibinfo{journal}{Phys. Rev. B}
  \textbf{\bibinfo{volume}{79}}, \bibinfo{pages}{214408}
  (\bibinfo{year}{2009}).

\bibitem[{\citenamefont{Frigo and Johnson}(2005)}]{FFTW05}
\bibinfo{author}{\bibfnamefont{M.}~\bibnamefont{Frigo}} \bibnamefont{and}
  \bibinfo{author}{\bibfnamefont{S.~G.} \bibnamefont{Johnson}},
  \bibinfo{journal}{Proceedings of the IEEE} \textbf{\bibinfo{volume}{93}},
  \bibinfo{pages}{216} (\bibinfo{year}{2005}), \bibinfo{note}{special issue on
  ``Program Generation, Optimization, and Platform Adaptation''}.

\bibitem[{\citenamefont{Liu and Kosloff}(1981)}]{LiuKosloff81}
\bibinfo{author}{\bibfnamefont{H.-P.} \bibnamefont{Liu}} \bibnamefont{and}
  \bibinfo{author}{\bibfnamefont{D.~D.} \bibnamefont{Kosloff}},
  \bibinfo{journal}{Geophys. J. Royal Astr. Soc.}
  \textbf{\bibinfo{volume}{67}}, \bibinfo{pages}{791} (\bibinfo{year}{1981}),
  ISSN \bibinfo{issn}{1365-246X}.

\end{thebibliography}

\end{document}